\documentclass[sigconf]{acmart}
\AtBeginDocument{%
  \providecommand\BibTeX{{%
    Bib\TeX}}}

\usepackage{acmart-taps}

\usepackage{tabularray} % Another table package for advanced features
\usepackage{tabularx} % Tables with flexible column width
\usepackage{booktabs}
\usepackage{longtable}
\usepackage{multirow} % Combine rows in tables
\usepackage{array} % Extended array and tabular environments
\newcolumntype{C}[1]{>{\centering\arraybackslash}p{#1}} % Centered version of p-column
\usepackage{capt-of}
\usepackage{wrapfig} % Wrap text around figures
\usepackage{float} % Improved control over float positioning

\usepackage{siunitx} % Consistent typesetting of units and numbers
\usepackage{etoolbox} % Conditional and robust commands
\usepackage{dcolumn} % Align decimal columns in tables
\usepackage{booktabs} % Professional-looking tables (rules, spacing)

\usepackage{cuted}
\usepackage{capt-of}
\let\liningnums\relax
\usepackage{fontspec}
\usepackage{fontawesome5}

\usepackage{xcolor} % Provides color support
\usepackage{soul} % Highlighting, underlining, and other text decorations
\soulregister{\cite}7 % Register cite commands so they won't break under soul
\soulregister{\citep}7
\soulregister{\citet}7
\soulregister{\ref}7
\soulregister{\pageref}7

\usepackage{framed}

\usepackage{xspace} % Insert space intelligently after macros

\usepackage{bm} % Bold math symbols
\newrobustcmd*{\bftabnum}{ %
  \bfseries
  \sisetup{output-decimal-marker={\textmd{.}}} %
}

\usepackage{pifont} 
\definecolor{oxfordblue}{rgb}{0.0, 0.13, 0.28}
\definecolor{harvardcrimson}{rgb}{0.79, 0.0, 0.09}
\definecolor{dartmouthgreen}{rgb}{0.05, 0.5, 0.06}
\definecolor{princetonorange}{rgb}{1.0, 0.56, 0.0}
\definecolor{yaleblue}{rgb}{0.06, 0.3, 0.57}
\definecolor{usccardinal}{rgb}{0.6, 0.0, 0.0}
\definecolor{uclablue}{rgb}{0.33, 0.41, 0.58}
\definecolor{msugreen}{rgb}{0.09, 0.27, 0.23}
\definecolor{cornellred}{rgb}{0.7, 0.11, 0.11}
\definecolor{pomegranate}{RGB}{192, 57, 43}
\definecolor{anti-pomegranate}{RGB}{43,178,192}
\definecolor{alizarin}{RGB}{231, 76, 60}
\definecolor{anti-belize}{RGB}{185, 41, 56}
\definecolor{belize}{RGB}{41, 128, 185}
\definecolor{sky}{RGB}{52, 152, 219}
\definecolor{green}{RGB}{22, 160, 133}
\definecolor{anti-green}{RGB}{160,22,118}
\definecolor{turquoise}{RGB}{26, 188, 156}
\definecolor{pumpkin}{RGB}{211, 84, 0}
\definecolor{anti-pumpkin}{RGB}{0,22,211}
\definecolor{carrot}{RGB}{230, 126, 34}
\definecolor{wisteria}{RGB}{142, 68, 173}
\definecolor{anti-wisteria}{RGB}{99,173,68}
\definecolor{amethyst}{RGB}{155, 89, 182}
\definecolor{nephritis}{RGB}{39, 174, 96}
\definecolor{anti-nephritis}{RGB}{174,39,117}
\definecolor{grey-bg}{RGB}{242,242,235}
\definecolor{light-bg}{RGB}{249,249,249}
\definecolor{extended-blue}{RGB}{59,130,246}
\definecolor{extended-red}{RGB}{239,68,68}
\definecolor{extended-orange}{RGB}{249,115,22}
\definecolor{extended-violet}{RGB}{99,102,241}
\definecolor{extended-green}{RGB}{16,185,129}

\newcommand{\revise}[1]{{\color{black} #1}}

\newcommand{\new}[1]{{\color{black} #1}}

\AtBeginDocument{ %
  \providecommand\BibTeX{{ %
    \normalfont B\kern-0.5em{\scshape i\kern-0.25em b}\kern-0.8em\TeX}}}

\acmSubmissionID{6429}

\copyrightyear{2026}
\acmYear{2026}
\setcopyright{cc}
\setcctype{by}
\acmConference[ASSETS '26]{The 28th International ACM SIGACCESS Conference on Computers and Accessibility}{October 25--28, 2026}{Vila Nova de Gaia, Portugal}
\acmBooktitle{The 28th International ACM SIGACCESS Conference on Computers and Accessibility (ASSETS '26), October 25--28, 2026, Vila Nova de Gaia, Portugal}
\acmDOI{10.1145/3797867.3828993}
\acmISBN{979-8-4007-2521-0/2026/10}

\begin{document}

\title[]{Beyond Harassment: Exploring the Harm Experienced by People with Disabilities in Social Virtual Reality}

\author{Xinran Adeline Li}

\orcid{0009-0006-9503-129X}
\affiliation{%
   \department{Department of Computer Science}
   \institution{Johns Hopkins University}
   \city{Baltimore}
   \state{Maryland}
   \country{USA}
}

\email{xli436@jhu.edu}
\renewcommand{\shortauthors}{Li et al.}

\author{Kexin Zhang}
\orcid{0009-0009-4078-8780}
\affiliation{%
   \department{Department of Computer Science}
   \institution{University of Wisconsin-Madison}
   \city{Madison}
   \state{Wisconsin}
   \country{USA}
}
\email{kzhang284@wisc.edu}

\author{Yuhang Zhao}
\orcid{0000-0003-3686-695X}
\affiliation{%
   \department{Department of Computer Science}
   \institution{University of Wisconsin-Madison}
   \city{Madison}
   \state{Wisconsin}
   \country{USA}
}
\email{yuhang.zhao@cs.wisc.edu}

\author{Yaxing Yao}
\orcid{0009-0008-7900-9265}
\affiliation{%
   \department{Department of Computer Science}
   \institution{Johns Hopkins University}
   \city{Baltimore}
   \state{Maryland}
   \country{USA}
}
\email{yaxing@jhu.edu}

\renewcommand{\shortauthors}{Li et al.}

\begin{abstract}
People with disabilities (PWD) are increasingly engaging in social virtual reality (VR) platforms, where immersive and embodied interactions can intensify negative experiences. While prior work has examined harassment in VR, little is known about the harms experienced by PWD and the perceived severity associated with different harassment and disability types. Unlike harassment, which represents behaviors, harm is more critical to designing effective protections, as it reflects the consequences and impact; the realism of VR and the vulnerability resulting from disability identity can further amplify such impact. To characterize and model harms for PWD, we conducted a literature review, followed by an online survey with 67 PWD to understand participants' harassment experiences and resulting harms in social VR. We identified 19 types of harm in 5 categories, and reported the severity perception of each type of harm. Finally, we analyzed our results from the critical disability theory perspective, summarized the uniqueness of harm in social VR, and discussed design implications for specialized safety mechanisms that mitigate harm for PWD.

\end{abstract}

% CCS XML, Keywords, etc.



% \begin{teaserfigure}
%  \includegraphics[width=\linewidth]{figures/teaser_figure.jpg}
%  \caption{\todo{The figure illustrates how users...}}
%  \Description{}
%  \label{fig:teaser_figure}
% \end{teaserfigure}

\maketitle

% Section I
\section{Introduction}
\label{sec:intro}

Social VR, as a virtual space for people to remotely meet, socialize, and interact in the form of avatars, has gained increased popularity~\cite{freeman2021body}. Unlike traditional social platforms, social VR uses full-body tracking, synchronous voice chat, and a wide range of interactive features to offer users a novel and immersive social experience~\cite{freeman2022disturbing}, bridging the gap between physical and digital presence. 

However, the realistic and embodied experience of social VR also introduces a new layer of complexity to social interactions~\cite{mei2021cakevr}, posing and even amplifying safety risks~\cite{blackwell2019harassment, kuhne2023direct, tian2021emotional}. Harassment, stigma, or other forms of negative experience in Social VR may feel more realistic than other online platforms, as the boundaries between virtual and real experiences become increasingly blurred~\cite{blackwell2019harassment}. 
\new{
To further study the impact of such experiences, we distinguish harassment from harm: harassment refers to threatening, harmful, or humiliating conduct based on race, color, national origin, sex, or disability~\cite{titleVI}; whereas harm refers to the damaging consequences produced by those behaviors, serving as the downstream impact of harmful experiences rather than the behavior itself~\cite{xiao_sensemaking_2022, ignatuschtschenko2016cyber}.
}

Such negative experiences can be further exacerbated for people with disabilities (PWD).
% The presence and experiences of people with disabilities (PWD) have begun to draw greater attention in social VR contexts. 
While PWD often choose to disclose their disabilities (e.g., through avatar design or 
% in various ways, such as using inclusive avatar designs, or 
in their profiles)~\cite{mack2023towards, zhang2022s} to improve social connections and
% . In particular, using avatars that reflect disability identity can also positively impact their social connections, 
foster meaningful conversations~\cite{gualano2024try, angerbauer2024part, mack2023towards}, such disclosure increases the risk of disability-targeted harassment in social VR~\cite{angerbauer2024part, zhang2023diary}. For example, visible disability signifiers on avatars may lead to bullying, rude comments, ableist language, and even direct insults~\cite{angerbauer2024part, zhang2023diary}. 

% found that participants using visible disability signifiers reported experiences of bullying, rude comments, and even direct insults. Similarly, Zhang et al.~\cite{zhang2023diary} showed that PWD using disability signifiers (DS) face varied harassment, including ableist language, being treated as inferior, mimicry through DS avatars, and unwanted interactions. Collectively, these studies demonstrate that PWD frequently experience targeted harassment in VR, drawing researchers’ attention to their social experiences in these environments.

Despite prior knowledge of harassment (\textit{the behaviors}), the harms (\textit{the impact}) on PWD caused by such virtual harassment have not been systematically investigated. 
% While prior work has identified the types of harassment PWD face in social VR, the specific harms and impacts remain unclear. 
Prior work has suggested several frameworks that captured different types of harms. For example, Scheuerman et al.~\cite{scheuerman2021framework} identified four types of harm caused by online content: physical harm, emotional harm, relational harm, and financial harm.
% \yaxing{@xinran, add an example}.
Yet, existing frameworks do not cover the unique experiences that PWD may have in social VR
% Prior work has proposed several harassment or harm frameworks
~\cite{al2013cyber, petruccelli2022taxonomy, ignatuschtschenko2016cyber, scheuerman2021framework, deldari2023investigation, webb2025user}.
In social VR, negative experiences can be exacerbated by the heightened sense of presence and embodiment~\cite{blackwell2019harassment, slater2009we, tham2018understanding, tschanter2025towards}. For example, Tschanter et al.~\cite{tschanter2025towards} found that harassment in social VR can evoke intense emotional reactions, with self-similar avatars amplifying the impact and making the experience more personal. 
Moreover, with the immersive VR dynamics intersecting with disability-related vulnerabilities, PWD may perceive harassment differently than others, leading to greater emotional harm, social withdrawal, or long-term psychological impacts~\cite{lee2022association, augustine2024role}. Quinn and Chaudoir~\cite{quinn2015living} demonstrated that individuals with stigmatized identities experience greater psychological distress when their identity is highly central within a culturally devalued context, highlighting their heightened vulnerability to perceived harms.
As a result, harms to PWD caused by harassment in social VR may feel more personal and complex than in other online spaces~\cite{blackwell2019harassment, slater2009we, tham2018understanding, tschanter2025towards}.

\new{
This leaves a critical gap in the literature. While prior work has largely focused on identifying forms of harassment, little attention has been paid to the harms caused by these experiences. The embodied nature of VR environments, combined with the unique aspects of PWD identity,
} 
highlights the urgent need for a systematic investigation of the potential harms that PWD may experience in social VR. In this paper, we ask the following research questions:

\begin{itemize}
    \item \textbf{RQ1}: What harms do PWD experience from harassment in social VR, and how are these harms connected to disability identity and the social VR environment?
    \item \textbf{RQ2}: How do PWD perceive the severity of these harms?
\end{itemize}

To answer these research questions, we conducted a survey study with 67 PWD participants to collect a wide range of harassment experiences in social VR and the associated harms. To ground the survey in prior work and inform the survey design, we first conducted a literature review to solicit a list of disability-centered harassment types and a list of harms that PWD may face in their lives. We asked participants to share the details of their past harassment incidents and explain the harms they experienced. In total, our participants reported 152 harassment incidents, the harms associated with them, and their perceived severity of the harms. By analyzing these harms, we identified 19 types of harm in 5 categories, including emotional harm, internalized harm, social harm, somatic recall harm, and sensational harm.
Many harm types are centered on participants’ disability identity,
e.g., disability identity concealment, which refers to PWD's hesitance or refusal to disclose their disability status even in social activities beyond VR after experiencing harassment.
\revise{In particular, some harm types are strongly connected with disability-related symptoms or tied to the unique characteristics of social VR. For example, somatic recall harm can reactivate or worsen the disability symptoms, and sensational harm may introduce phantom sensations. In addition, some subtypes may appear similar to harms identified in prior work centered on the general population, but they may reinforce feelings of abnormality and loss of belonging for PWD.}

This paper makes the following contributions. First, we present a taxonomy of harms associated with PWD's harassment experiences in social VR. This taxonomy includes 19 types of harms spanning across 5 categories. 
Second, we report the perceived severity of each type of harm. 
Third, we rethink the harm from a critical disability theory perspective, discuss the unique forms of harm in social VR, and provide concrete design implications for preventing harm in social VR at different stages of harassment incidents.

% Research Question
% RQ1: What types of harm are perceived by PWD resulting from harassment in social VR? 
% RQ2: What harms are unique to disability and social VR?
% RQ3: What are the factors that shape PWD's perceived severity of the harm?
% RQ4: What is the connection between the each harm type and harassment type?

\section{Related Work}
\label{sec:relwork}

% \xinran{TODO: reframe the sentence, delete too complex}
\subsection{Stigma and Harassment Faced by PWD}

PWD often face stigma and discrimination that harm their mental health and social well-being~\cite{putnam2003health, tough2017social}. They may be seen as less capable, left out of group activities, or targeted with bullying and verbal abuse~\cite{heung2022nothing, heung2024vulnerable, equality2011hidden}. Such negative treatment can happen in schools, workplaces, and even within families~\cite{equality2011hidden, lindsay2023ableism, holzbauer2010typology, holzbauer1996disability}. Using assistive devices like wheelchairs, canes, or communication aids can also draw unwanted attention, causing others to see difference instead of inclusion~\cite{shinohara2011shadow, howard2022exploring}. Over time, repeated harassment can lead to stronger feelings of loneliness~\cite{deacon2018loss}. Even though awareness of disability rights has grown, many PWD still struggle to be treated fairly in everyday life~\cite{equality2011hidden, holzbauer1996disability}, and this ongoing marginalization can contribute to feelings of loneliness, making online spaces places they actively turn to not only for social connection but also for enjoyment and a sense of autonomy~\cite{ma2015mental, viluckiene2015relationship}.

Online platforms, especially social media, provide an alternative space for PWD to build connections and engage with others beyond physical limitations. However, previous research shows that PWD still face cyberbullying and online discrimination in these environments~\cite{heung2022nothing, heung2024vulnerable, alhaboby2016language, sannon2023disability, kowalski2016cyberbullying, kowalski2018cyberbullying, burch2018you}. 
% Alhaboby et al.~\cite{alhaboby2016language} surveyed 115 people, including 19 with disabilities, and found that disabled participants often faced cyberharassment involving disability slurs, identity misrepresentation, which raised concerns about family safety and led to negative psychological impacts. 
Heung et al.~\cite{heung2022nothing} interviewed 20 disabled individuals and identified 12 types of ableist microaggressions on social media, such as infantilization, accusations of faking disability, and exclusion through inaccessible content; participants reported harms including loss of self-confidence and 
experiencing traumatized memories. Later work by Heung et al.~\cite{heung2024vulnerable} and Sannon et al.~\cite{sannon2023disability} found that disabled content creators frequently faced harassment such as hateful messages, sexual harassment, and targeted suppression of disability content, leading to emotional distress and severe mental health impacts.

% For people with learning or intellectual disability, the increased desire for social interaction and tendency to trust strangers online can make them more vulnerable to online harassment~\cite{chadwick2019online, normand2016cybervictimization}. With difficulty recognizing deception, they face higher risks of encountering harmful content, financial scams, and sexual exploitation~\cite{chiner2017internet, holmes2014experiences}. Especially for young people~\cite{normand2016cybervictimization}, the lack of awareness and training in recognizing online risks can lead to more severe consequences. Such experiences of harassment and exploitation have been linked to psychological harms including anxiety, loneliness, and depression~\cite{chadwick2019online}.

Although prior work has examined stigma, harassment, and related harms faced by PWD in both offline contexts and online platforms, little research has focused on the specific harms they experience in social VR.

\subsection{Harassment in Social VR}

%% a para about immersive vr
In recent years, VR has become increasingly popular,  which has led to the emergence of social VR~\cite{freeman2021body, zamanifard2019togetherness}. Compared to traditional PC-based virtual worlds, social VR provides unique features of embodied avatars, simulated face-to-face real-time interactions via full-body tracking, and various virtual content, enhancing the virtual social environment to resemble the real world~\cite{maloney2020anonymity, kolesnichenko2019understanding, zhang2023diary, mcveigh2019shaping, zheng2023understanding}. However, these unique features can also lead to potential risks of more complex forms of harassment, compared with traditional text- or voice-based online harassment~\cite{blackwell2017classification}. Furthermore, due to the realistic and immersive nature of social VR, harassment that occurs in social VR can be even more immersive and destructive~\cite{freeman2022disturbing, blackwell2019harassment}, leading to the complexity of negative impacts~\cite{kuhne2023direct, tian2021emotional}.

Previous studies have explored harassment and privacy concerns within social VR environments~\cite{blackwell2019harassment, freeman2022disturbing,zheng2023understanding, gray2024content, shriram2017all, maloney2020anonymity, boping2024sexual}. Shriram and Schwartz~\cite{shriram2017all} found that harassment is common in social VR, with 21 of 99 men and 2 of 7 women reporting harassment, and 42\% of users witnessing it. Moreover, researchers~\cite{blackwell2019harassment, freeman2022disturbing, gray2024content} identified three types of harassment: verbal, physical, and environmental harassment, noting that VR’s immersive features can intensify harm. 

% Freeman et al.~\cite{freeman2022disturbing} highlighted four characteristics of VR harassment, including disruptive physical behaviors, forced attention via voice chat, and personal space invasion. Similarly, Gray et al.~\cite{gray2024content} showed that harassment is not limited to physical contact, as embodied presence like standing too close, blocking movement, or suggestive gestures can also cause discomfort and harm.

Individuals from underrepresented groups (e.g., LGBTQ, people with disabilities, and racial minority groups) may face a higher risk of harassment in social VR~\cite{heung2024vulnerable, freeman2022disturbing, freeman2021body, maloney2020complicated, blackwell2019harassment2, zhang2023diary, connor2023consensual}. Prior work has shown that women, LGBTQ+ individuals, and ethnic minority participants often experience more disruptive forms of harassment due to the visibility of their gender, sexual orientation, or ethnic identity in avatar design and voice~\cite{freeman2021body, freeman2022disturbing}. However, recent research highlights that PWD face unique forms of targeted harassment in these spaces. For example, Angerbauer et al.~\cite{angerbauer2024part} found that participants using avatars with visible disability signifiers, such as wheelchairs or canes, were more likely to encounter bullying, rude comments, and direct insults, which negatively affected their experiences. Similarly, Zhang et al.~\cite{zhang2023diary} reported that PWD who used disability signifiers encountered repeated ableist language, unwanted interaction with disability signifiers, and other physical harassment directed toward their avatars. 

Together, these studies reveal that PWD who disclose their disability face higher risks of being harassed. Despite growing evidence on harassment in social VR, little work has focused on the harms associated with these experiences.

\subsection{Harm Framework for Online Space}
Prior research has developed concrete frameworks to analyze online harassment~\cite{thomas2021sok,zheng2023understanding, pater2016characterizations, banko2020unified, al2013cyber, maarten2006class}. Thomas et al.~\cite{thomas2021sok} proposed a taxonomy of online attacks, while Zheng et al.~\cite{zheng2023understanding} outlined risks in social VR such as virtual violence, abuse, and sexual harassment. These frameworks provide background for our study of harassment and harm in VR. However, they focus on companies and infrastructures rather than on individuals.

While many frameworks have examined harassment, few have developed systematic taxonomies for categorizing online harms~\cite{scheuerman2021framework, al2013cyber, ignatuschtschenko2016cyber, petruccelli2022taxonomy, webb2025user, klonick2015new, agrafiotis2018taxonomy}. 
Prior work on cyber harm has proposed conceptual and taxonomic frameworks for understanding and measuring different kinds of impacts, such as physical, economic, psychological, and reputational harms~\cite{ignatuschtschenko2016cyber, agrafiotis2018taxonomy, klonick2015new, al2013cyber}. At the individual level, other work categorizes harms such as social and cultural, psychological, and financial harms, including in cyberbullying and metaverse contexts~\cite{al2013cyber, webb2025user}. Research on online harmful content further develops taxonomies and severity models that link content types to downstream harms and highlight factors like intent, vulnerability, and scale~\cite{banko2020unified, thomas2021sok, scheuerman2021framework}.

% For cyber harm, Agrafiotis et al.~\cite{ignatuschtschenko2016cyber} offered a conceptual framework for understanding and measuring cyber harm, identifying six types of harm and five guiding questions to assess the harm. At the individual level, there are also frameworks categorizing different harms such as social and cultural, psychological, financial harms~\cite{al2013cyber, webb2025user}. For online harmful content, Scheuerman et al.~\cite{scheuerman2021framework} then developed a harm framework based on interviews and card-sorting with 52 participants, identifying four types of harms and presenting factors such as intent, vulnerability, and scale that shape the severity of harm.

Although prior work has provided frameworks for harassment experiences and harm in online spaces, little work has specifically focused on the experiences of PWD. 
\revise{Existing harm taxonomies only summarize the harms in broad categories such as psychological harm, social harm, and reputational harm; they do not include deeper connections between these harms and disability identities and disability-related symptoms. They also do not include harms that are unique to the social VR context, where some harms can be amplified by the embodiment feature of social VR. }
Based on the unique nature of social VR and the vulnerabilities of PWD, a structured investigation centered on their harassment experiences and harms in the social VR context is needed. Our study aims to address this gap.

\section{Method}
\label{sec:method}
 % 1. Do a literature review to identify the types of harassment PWD face in social VR and online space

 % 2. Design a survey to ask about harassment experience and perceived harm

% To answer our research questions, we conducted an online survey with 67 participants. 
% who reported 152 number of harassment instances.
% after screening the low-quality ones. 
% We also conduct a literature review to initially identify the types of harassment and harm in our survey. 
To provide a structured way for our participants to share their harassment experiences, we first conducted a literature review to identify the types of disability-related harassment reported by prior work. Informed by the literature review, we then conducted a survey to collect participants' experiences and perceptions of social VR harassment and harms. In this section, we present the study details.
% identifying the harassment types faced by PWD and the associated harms, the survey protocol, participants recruitment, and data analysis. 
This study was approved by our university IRB. 
% We also implemented strict data management rules to ensure the ethical conduct of our research (e.g., we collected and stored survey data in our university-approved cloud services and only allowed access to researchers involved in this project).

\subsection{Literature Review}
\label{subsec:identifying-harassment-harm}

The goal of the literature review is to solicit the types of disability-related harassment that PWD may experience. 
% To better help participants share their harassment incidents in social VR, we first conducted a systematic literature review to identify the common
% To understand the harms of harassment faced by PWD and inform our survey design, we conducted a literature review to identify and categorize existing 
% types of disability-related harassment.
% We aim to cover as many types of harassment as possible in the survey to ensure its comprehensiveness. 
% This process ensured that our survey captured a wide range of harassment experiences from multiple perspectives. 
We searched on Google Scholar and other key academic databases in HCI and AR/VR (i.e., ACM CHI, CSCW, ASSETS, and IEEE VR) using combinations of several keywords, including `online harassment', `VR harassment experience', `people with disabilities', `hate and harassment', `disability harassment', and `social virtual reality'. We manually filtered the search results by reviewing the content of each paper and removing irrelevant (e.g., not centered on disability) and repetitive papers. Given the limited work that focuses on PWD's harassment experience specifically in social VR, as well as the fact that VR commonly simulates real-world experiences, we also included PWD's harassment incidents in the real world and in online spaces to capture the broader context more generally. 

In total, we identified 32 papers related to PWD's harassment experiences. We extracted all harassment types described in these papers, compared how each work defined and contextualized harassment, and then consolidated overlapping descriptions to form a unified definition for our analysis.
% \yaxing{@xinran, add some analysis details}
In the end, we identified 8 types of harassment, including verbal harassment (e.g., insults, slurs, and mockery centered on disability, social exclusion, and asking invasive personal questions), physical harassment (e.g., aggressive physical interactions in social VR), and special harassment types that only target PWD (e.g., intentionally imitating disability-related traits in social VR).
We detailed their definitions in Table~\ref{tab:harassment_def} in Appendix~\ref{tab:harassment-definitions}.

\subsection{Survey Protocol}

 % \xinran {TODO: add example questions, refine to paragraph, add flow chart?}
 
 % Next, we introduce our survey protocol. We designed the survey to investigate the details of PWD's harassment experience and the associated harm in social VR. 
Our survey includes three sections as listed below. All questions were mandatory.
 The complete survey protocol is available in the Appendix~\ref{tab:survey_protocol}.
 
 % consisted of four parts, demographic, social VR experience, general harassment experience, and detailed descriptions on specific harassment types. 
\textbf{Participants' background.} 
We first asked participants for their demographic information and their general experiences with social VR, including how frequently they used it, their total usage time, the platforms they had used, and how they disclosed their disability in those spaces.

\textbf{Harassment experiences.} 
In this part, we first asked participants to provide their own definitions of harassment. We then asked participants to report their harassment experiences. To do so, participants were first asked to select all applicable harassment categories from a list of eight types we identified in the previous step (with definitions provided) or to add a new category if their experience was not covered. Then, the survey expanded each selected type into a follow-up section in which participants were asked to provide detailed information, including where the incident occurred, on which platform, the number of offenders involved, their relationship to the offenders, and a description of the incident.

\textbf{Harm experiences.} 
For each harassment incident reported by the participant, we asked them follow-up questions to understand the harms the harassment may have caused. 
We first presented the four harm types (inspired by the harm frameworks of Agrafiotis et al.~\cite{agrafiotis2018taxonomy}, Scheuerman et al.~\cite{scheuerman2021framework}, and Petruccelli et al.~\cite{petruccelli2022taxonomy}) and asked whether they had experienced any of these harms. Similarly, participants could also report new types of harms if their experiences were not covered by the predefined categories. For each harm type marked ``Yes,'' we asked follow-up questions, including a detailed description of the harm, a severity rating on a five-point scale (ranging from ``Not severe at all'' to ``Very severe'') and an explanation of the severity rating. 
If participants reported a harassment incident without any harm, we asked them to explain why. After participants went through all harm experiences, we further asked them to compare the harassment incidents and the associated harms in social VR with those in the real world.

\subsection{Participant Recruitment and Eligibility}
We implemented our survey on Qualtrics. To distribute our survey, we reached out to non-profit organizations that work with people with various disabilities (e.g., the National Federation of the Blind, Invisible Disability Association, and Amputee Coalition)
% \yuhang{add more disability organizations that focus on other disabilities; blind people are not typical social VR users}) 
and asked them to post our study to their mailing lists or on their own social media platforms (e.g., Facebook, Instagram). Interested participants first completed a screening survey with their age, disability
conditions, and general social VR and harassment experiences. Participants were eligible if they (1) were over 18 years old, (2) self-identified as having disabilities, (3) had disclosed their disability, and (4) self-identified as having experienced harassment in social VR. 
\new{
To validate participants' eligibility, we asked them several questions about their disability status in the screening survey, such as ``Do you have any disabilities or chronic conditions?'' ``What disabilities do you have? Please select all that apply,'' and ``Please briefly describe how long you have had the disabilities selected above and how they affect your daily life.'' These responses were used to confirm that participants met the recruitment criteria.
}
In total, we recruited 67 participants with diverse disabilities. The average time of survey completion was 50 minutes. Upon completion and the validation of their responses, each participant received \$10 as compensation.

\subsection{Survey Development and Pilot Study}
To evaluate the survey flow, structure, and screening mechanism, we conducted a pilot study with 12 participants. Based on the feedback and observations from the pilot, we refined the survey by adjusting the flow to make it easier to navigate, revising question wording for clarity, and correcting errors in the branching logic. These revisions resulted in the final version of the survey used for the formal study.

\subsection{Data Analysis}

% \subsubsection{Data Cleaning}
\textbf{Data cleaning.} 
We started the data analysis by first cleaning the data. \revise{We used multiple methods to prevent fraud, scams, and low-quality responses. First, we enabled bot detection and duplicate submission prevention features in Qualtrics. Before sending out the main survey, we carefully selected participants based on their IP addresses, and we filtered out IP addresses that were not in the US and responses from the exact same IP address. We also removed responses with a completion time of less than 10 minutes (average completion time was 50 minutes). Since we collected the demographic data in the screening survey, we also removed responses that contained invalid or inconsistent demographic information (e.g., fake or inconsistent disability descriptions). After we collected the responses to the main survey, we checked each participant’s disability type and demographics again to ensure consistency across both surveys.}
% 2) checked participants' IP addresses and removed inconsistent, duplicated, or non-US IP addresses; 3) removed responses with a completion time of less than 10 minutes (average completion time was 50 minutes); 4) manually reviewed and removed low-quality answers (e.g., repeated answers for all questions); and 5) removed responses from the main survey if demographic data did not match the screening survey.
% Participants first completed a short screening survey about their disability and demographics. Two researchers agreed on criteria for identifying invalid responses (e.g., fake or inconsistent disability descriptions), and one researcher applied these criteria to remove suspicious entries. After the main survey, we checked each participant’s disability type and demographics again to ensure consistency across both surveys. 

After the data cleaning, we then analyzed the harassment and harm descriptions, assigned each incident to our predefined categories, and merged repeated incidents from the same participant when they described the same category.

\textbf{Qualitative Analysis.}
For the open-ended questions, \revise{we followed Reflexive Thematic Analysis~\cite{braun2006using, braun2022thematic} to identify patterns and themes in the data. Since our goal was to understand PWD's harm experiences in social VR and how those harms were connected to disability-related contexts, we used an inductive, bottom-up approach to generate harm types from participants' responses.}
Two researchers first coded a 21\% subset of responses independently at the sentence level, then discussed and reconciled their codes to create an initial codebook. Using this codebook, one researcher recoded the full dataset and made revisions as needed. All authors then reviewed the codes, resolved disagreements, and refined the codebook until full agreement was reached, producing separate codebooks for different survey questions. \revise{The higher-level themes are framed as harm categories that map onto our research questions, and the sub-themes capture the more specific characteristics, contexts, and experiences within each harm category.}

For harm category generation and harassment experience analysis, although we provided predefined harm and harassment categories, researchers conducted the thematic analysis solely based on participants’ original descriptions rather than on the predefined categories. After finalizing the codebook, we recategorized all codes into the corresponding harm and harassment categories. The final codebook contained 37 codes in total. We then used an affinity diagram to classify all codes into broader themes according to our research questions. 
% Because the coding was collaboratively conducted in multiple rounds with complete agreement achieved, intercoder reliability was not necessary~\cite{mcdonald2019reliability}.

% For open-ended questions, we used the method of thematic analysis~\cite{braun2006using, braun2013successful} to identify repetitive patterns and themes from our data. Two researchers first coded a 21\% subset of the data independently at the sentence level. They then discussed and reconciled their codes to resolve differences, developing an initial codebook upon reaching agreement. With this initial codebook, one researcher then re-coded the entire data set, making alterations as required. Upon completion, all authors reviewed the codes together, discussed and resolved any disagreements, and refined the codebook as needed until a full agreement was reached. We form separate codebooks for different survey questions .\revise{The final codebook contained 37 codes in total.} Using the final codebook, we conducted a thematic analysis, classifying all codes into themes according to our research questions. Because the coding was collaboratively conducted in multiple rounds with complete agreement achieved, intercoder reliability was not necessary~\cite{mcdonald2019reliability}.

\textbf{Quantitative Analysis.} 
\revise{To complement our qualitative coding, we conducted quantitative analyses to identify patterns in harassment and harm. Our quantitative analysis mainly consisted of descriptive statistics. We summarized participant demographics with counts and percentages, and after identifying harm types through qualitative coding, we calculated severity rating statistics for each harm type, including counts, minimum, maximum, mean, and variance for each type.}

\subsection{Ethics Considerations and Positionality}

We obtained IRB approval from our university for all human participant research described in this manuscript. This study examines harassment and harm experienced by people with disabilities in social VR, which may involve recalling distressing events. To reduce discomfort, participants were informed that they could skip any questions or withdraw at any time, and the survey language was reviewed to ensure sensitivity. All responses were anonymous, and no personally identifiable information was collected. Because some questions ask about past harassment, the research team follows applicable state laws and will report suspected abuse or risk of harm involving a child or dependent adult.

\subsection{Positionality Statement}
\revise{We acknowledge the importance of positionality when working with marginalized populations~\cite{liang2021embracing}. All authors have rich experience conducting research with the disability community. We reviewed the survey protocol to ensure clarity, inclusiveness, and sensitivity. We aim to accurately understand how people with disabilities experience and define harassment and harm in social VR.}
\section{Results}
\label{sec:results}

To answer our research questions, we present five broad categories of harm. For each category, we describe the definitions of sub-types and explain them in detail. We further report the severity perception of each harm type.

\subsection{Demographic Information}

\begin{table*}[ht]
\centering
% \tagpdfsetup{table/header-rows={1,8}}
\renewcommand{\arraystretch}{1.2}
\footnotesize 

\begin{tabular}{
p{2.60cm}p{0.75cm}
p{2.70cm}p{0.75cm}
p{3.60cm}p{0.75cm}
p{3.05cm}p{0.75cm}
}
\toprule
\multicolumn{2}{c}{\textbf{Gender}} &
\multicolumn{2}{c}{\textbf{Age}} &
\multicolumn{2}{c}{\textbf{Ethnicity}} &
\multicolumn{2}{c}{\textbf{Social VR Usage}} \\
\cmidrule(lr){1-2}
\cmidrule(lr){3-4}
\cmidrule(lr){5-6}
\cmidrule(lr){7-8}
Female & 48.5\% &
25 to 34 & 54.5\% &
White & 50.0\% &
More than 2 years & 47.0\% \\
Male & 39.4\% &
35 to 44 & 21.2\% &
Black or African American & 33.3\% &
1 to 2 years & 30.3\% \\
Non-Binary & 9.1\% &
18 to 24 & 16.7\% &
Asian & 9.1\% &
1 to 6 months & 15.2\% \\
Prefer not to say & 3.0\% &
55 to 64 & 4.5\% &
American Indian or Alaska Native & 3.0\% &
7 to 12 months & 7.6\% \\
& &
45 to 54 & 1.5\% &
Prefer not to say & 3.0\% &
& \\
& &
65 or over & 1.5\% &
Hispanic or Latino & 1.5\% &
& \\
\midrule
\multicolumn{4}{c}{\textbf{Disability Type}} &
\multicolumn{2}{c}{\textbf{Disability Disclosure Method}} &
\multicolumn{2}{c}{\textbf{Social VR Frequency}} \\
\cmidrule(lr){1-4}
\cmidrule(lr){5-6}
\cmidrule(lr){7-8}
Autism & 21.4\% &
Health-related disability & 6.0\% &
Verbal conversation with others & 41.1\% &
Frequently & 43.9\% \\
Mobility-related disability & 21.4\% &
Deaf or Hard of Hearing & 2.4\% &
Profile information or bio & 27.4\% &
Occasionally & 31.8\% \\
Mental health condition & 10.7\% &
Multiple sclerosis & 1.2\% &
Through avatar appearance & 26.3\% &
Very frequently & 19.7\% \\
Speech-related disability & 9.5\% &
Neurological & 1.2\% &
Floating name tag or status & 5.3\% &
Rarely & 3.0\% \\
Blind or low vision & 8.3\% &
Muscular dystrophy & 1.2\% &
& &
Very rarely & 1.5\% \\
Learning disability & 8.3\% &
Neurological disorder & 1.2\% &
& &
& \\
Attention deficit & 7.1\% &
& &
& &
& \\
\bottomrule
\end{tabular}%
\vspace{4pt}
\caption{Participants’ demographic information.}
\label{tab:demographics}
\end{table*}

We recruited 67 participants with diverse backgrounds in terms of gender, age, and ethnicity. All participants were active social VR users with varying levels of experience and engagement frequency. Participants also represented a range of disability types. They also used different methods to disclose their disabilities, such as through conversation, profile information, or avatar appearance. The detailed demographic information is summarized in Table~\ref{tab:demographics}.

\subsection{PWD's Harassment Incidents}

Our participants reported a total of 152 harassment experiences. Most incidents fell under the 8 categories we identified through the systematic literature review, including accusations of faking disability (N=11), disability-centered insults and mockery (N=43), invasive personal questions (N=14), treating disability as inability (N=13), physical harassment (N=16), sexual harassment (N=11), social exclusion (N=28), and mimicking disability (N=12).

Although relatively few, participants also reported two new types of harassment that have not appeared in prior work, i.e., intentionally triggering disability symptoms (N=3) and indirect harassment (N=1). Intentionally triggering disability symptoms refers to deliberate actions in social VR, after a disability has been disclosed, where harassers intentionally behave or manipulate the environment to provoke specific symptoms of a person’s disability. As P2 (male, health-related disability, mental health condition) revealed, \textit{``The person picked their nose when I shared my Mysophobia with them and in particular nose picking... When I shared that I have an extreme fear of people who pick their nose, they picked their nose and laughed.''} Indirect harassment refers to situations where people with disabilities witness or overhear harassment in social VR, such as insults, mocking, or mimicking, which can still cause harm even without direct engagement. As P54 (male, mobility-related disability) described, \textit{``Suddenly, I overheard a group making fun of someone with a visible disability. They were laughing and using slurs, thinking it was funny...It made me feel angry and upset, knowing that some people think it's okay to belittle others.''}

Next, we will focus on the harms associated with the harassment incidents, and examine how these experiences affected PWD.

\subsection{Five Categories of Harm}

We identified 5 high-level categories of harm triggered by disability-centered harassment, including emotional harm, internalized harm, social harm, somatic recall harm, and sensational harm. Each category further contains several subtypes of harm. We summarize them in  Table~\ref{tab:harm_def}.
% For each subtype, we also reported participants' severity ratings on a five-point scale (ranging from Not severe at all to Very severe), and factors that influence their ratings.
Although some harm types align with prior work, many harms (bolded in Table~\ref{tab:harm_def}) are either distinctive or more intense among PWD, showing the unique connections between the harms and disability-related harassment and contexts.

\begin{table*}[!t]
\centering
\footnotesize
\renewcommand{\arraystretch}{1.2}

%% with bolded items indicating harms that are distinctive among people with disabilities or show stronger connections to disability related contexts.}
% \tagpdfsetup{table/header-rows={1}}
\begin{tabular}{@{}p{1.5cm}p{4cm}p{4.0cm}p{6.5cm}@{}}

\toprule
\textbf{Harm Type} & \textbf{Definition} & \textbf{Sub-Type} & \textbf{Definition} \\
\midrule

Emotional Harm 
& The negative emotional impact caused by harassment, such as anger, anxiety, or humiliation. 
& Anger Triggered by Harassment 
& Participants' expressions of feeling angry or annoyed in response to harassment incidents. \\
\cmidrule(lr){3-4}
& 
& Anxiety and Emotional Strain from Harassment 
& A heightened emotional state marked by anxiety, stress, unease, or fear. \\
\cmidrule(lr){3-4}
& 
& Emotional Sadness after Harassment 
& A state of low mood or emotional downturn, characterized by sadness, disappointment, hurt, depression, or emotional exhaustion. \\
\cmidrule(lr){3-4}
& 
& \textbf{Disrespect and Humiliation from Mockery} 
& Feelings of being disrespected, demeaned, or embarrassed as a result of harassment. \\
\cmidrule(lr){3-4}
& 
& \textbf{Emotional Isolation due to Exclusion} 
& The emotional experience of feeling alone, excluded, or disconnected due to harassment. \\
\midrule

Internalized Harm 
& A form of psychological harm that extends beyond negative emotions captured under emotional harm. 
& \textbf{Cognitive Intrusion} 
& Persistent harassment-related thoughts, memories, or mental images that disrupt cognitive balance. \\
\cmidrule(lr){3-4}
& 
& \textbf{Erosion of Self-Worth} 
& A diminished sense of self-confidence and personal value caused by harassment. \\
\cmidrule(lr){3-4}
& 
& Harmful Coping Through Substances 
& Reliance on alcohol or prescription medication as a way to cope with harassment in social VR. \\
\cmidrule(lr){3-4}
& 
& Difficulties with Sleep 
& Trouble falling asleep, staying asleep, or experiencing nightmares following harassment incidents. \\
\cmidrule(lr){3-4}
& 
& Dissociation and Focus Disruption 
& Feeling detached from oneself, emotionally numb, or unable to concentrate and maintain daily routines. \\
\cmidrule(lr){3-4}
& 
& Self-Harm Desire 
& Thoughts of harming oneself. \\
\midrule

Social Harm 
& The negative impact on social connections, relationships, and perceived safety in social environments. 
& \textbf{Social VR Avoidance} 
& Reducing or avoiding social interaction in virtual social spaces after harassment. \\
\cmidrule(lr){3-4}
& 
& \textbf{Offline Social Withdrawal} 
& Reduced willingness to form new connections, make friends, or participate in real-world social activities after harassment in VR. \\
\cmidrule(lr){3-4}
& 
& \textbf{Disability Identity Concealment} 
& Reluctance or refusal to disclose disability status in social activities after harassment. \\
\cmidrule(lr){3-4}
& 
& \textbf{Erosion of Social Trust} 
& A diminished sense of security, belonging, and confidence in social VR or real world social environments. \\

\midrule

\textbf{Somatic Recall Harm} 
& Situations where the body ``remembers'' and reactivates symptoms in response to harassment. 
& \textbf{Disability-Related Symptoms Reactivation} 
& Harassment that activates or worsens disability-related symptoms, such as gastritis flare-ups, autistic responses, neurological issues, panic attacks, or other condition-specific symptoms. \\
\cmidrule(lr){3-4}
& 
& \textbf{Trauma Formation and Re-Triggering} 
& Harassment incidents that trigger past trauma or create new trauma that may persist. \\
\midrule

Sensational Harm 
& The negative impact of harassment incidents via phantom sensations in immersive VR, where the brain generates physical sensations without physical stimuli. 
& \textbf{Phantom Sensations} 
& Unreal but vivid sensory experiences closely linked to disability-targeted harassment. \\
\cmidrule(lr){3-4}
& 
& Embodied Reactions 
& Physical responses and discomfort during or after harassment in VR, from temporary sensations to more intense pain. \\
\bottomrule
\end{tabular}

\vspace{4pt}
\caption{\revise{Summary of harm categories and subtypes. Bolded items indicate harms that are distinctive among people with disabilities, or show stronger connections and are interpreted differently in disability-related contexts.}}
\label{tab:harm_def}

\end{table*}

% \begin{table*}[!t]
% \centering
% \footnotesize
% \renewcommand{\arraystretch}{1.2}
% \caption{Harm types and definitions (somatic recall and sensational harm).}
% \label{tab:harm_def_part2}
% \begin{tabular}{@{}p{1.8cm}p{4.0cm}p{3.2cm}p{5.0cm}@{}}
% \toprule
% \textbf{Harm Type} & \textbf{Definition} & \textbf{Sub-Type} & \textbf{Definition} \\
% \midrule

% \textbf{Somatic Recall Harm} 
% & Situations where the body ``remembers'' and reactivates symptoms in response to harassment. 
% & \textbf{Disability-Related Symptoms Reactivation} 
% & Harassment that activates or worsens disability-related symptoms, such as gastritis flare ups, autistic responses, neurological issues, panic attacks, or other condition-specific symptoms. \\
% \cmidrule(lr){3-4}
% & 
% & \textbf{Trauma Formation and Re-Triggering} 
% & Harassment incidents that trigger past trauma or create new trauma that may persist. \\
% \midrule

% Sensational Harm 
% & The negative impact of harassment incidents via phantom sensations in immersive VR, where the brain generates physical sensations without physical stimuli. 
% & \textbf{Phantom Sensations} 
% & Unreal but vivid sensory experiences closely linked to disability-targeted harassment. \\
% \cmidrule(lr){3-4}
% & 
% & Embodied Reactions 
% & Physical responses and discomfort during or after harassment in VR, from temporary sensations to more intense pain. \\
% \bottomrule
% \end{tabular}
% \end{table*}

\subsubsection{Category \#1: Emotional Harm}
Emotional harm refers to the negative emotional impact caused by harassment, such as feelings of anger, anxiety, or humiliation. Emotional harm affects PWD's emotional condition without necessarily resulting in long-term psychological effects.
% It is consistently centered on the harassment itself. 
This type of harm includes feelings of anger, anxiety, or hurt, often resulting from harassment targeting disability identity or unfair treatment. These harms primarily originate from the harassment incident itself and can bring various negative emotions to PWD. Moreover, they often manifest through language that reflects negative emotional effects (e.g., \textit{`I felt scared'}, \textit{`I was deeply hurt'}), indicating that the PWD is dealing with negative feelings, rather than experiencing deeper cognitive disruption.
% \yaxing{it is useful to also say where does emotional harm typically come from.}

\textbf{Anger Triggered by Harassment.}
Anger captures participants' expressions of feeling angry or annoyed in response to harassment incidents. 
% Participants described how the harmful interactions made them feel angry, annoyed, or mad. 
Participants (N=29) often expressed being angry, mad, or annoyed because of harassment targeting their disability identity. 
For example, P37 (male, deaf or hard of hearing) mentioned the accusation targeted at his disability triggered anger,
\textit{``During the experience, I felt angered by the fact that someone thinks that I am lying about my disability.''} 
Some participants also mentioned that being labeled casually, treated unfairly, or misunderstood because of their disability can lead to strong anger. 
As P45 (male, mental health condition and mobility-related disability) described the experience of being labeled, \textit{``You didn't say anything at all, but the other person casually labeled you and even made fun of you, saying that you `can't do anything for sure' or `just lie around playing games all day long' and so on.''} 

In some cases, bystanders' reactions can also be a source of anger, indicating that harm is not solely caused by the harasser but can also arise from the actions of those observing the incident targeted at disability. P50 (prefer not to say, blind or low vision) explained that the lack of action from the bystanders led to the feeling of anger, \textit{``Someone then started talking about how people with disabilities should not be allowed to reproduce, as they are useless...The others did not do anything and actually agreed with the bullying... I am angry that others did nothing to help and sided with the bullies.''} 
% \yaxing{we need a bit of explanation for each quote, see above for an example}

\textbf{Anxiety and Emotional Strain from Harassment.}
This harm type refers to a heightened emotional state marked by anxiety, stress, unease, or fear. Participants (N=30) reported emotional strain, such as feeling anxious, nervous, scared, or stressed after or during the harassment. 
For harm related to disability, P10 (female, attention deficit, blind or low vision and mental health condition) mentioned being anxious because others did not respect her disability boundaries, \textit{``I just felt anxious in the moment because someone was not respecting my boundary about my disability.''} P2 (male, health-related disability and mental health condition) also offered a special case where the harasser intentionally triggered Mysophobia, which led to increased anxiety, 
% \yaxing{is mysophobia considered as a type of disability?}
\textit{``The person picked their nose when I shared my Mysophobia with them and in particular nose picking...Seeing nose picking causes me great anxiety, and this increased my anxiety, and I had a hard time not thinking of it for a while.''}

\textbf{Emotional Sadness after Harassment.}
This harm refers to a state of low mood or emotional downturn, often characterized by feelings of sadness, disappointment, hurt, depression, or emotional exhaustion. Our participants (N=37) often mentioned feeling sad or depressed due to harassment that targeted their disabilities. As P50 (prefer not to say, blind or low vision) expressed sadness when being treated as incapable because of their disability, \textit{``I truly felt sad when someone said that they would not want to live if they were blind.''} 
\revise{P38 (female, deaf or hard of hearing and mobility-related disability) also felt depressed after being insulted because of disability, ``He kept referring to me as crippled because I was obviously in a wheelchair …. I noticed I started to slip and give in to depression because I couldn't handle it anymore.''}
Beyond the sadness caused by the harassment itself, P2 (male, health-related disability and mental health condition) also shared about feeling sadder after others discussed disability in front of him, \textit{``It makes me sadder than I already am when someone laughs at the details of what I share about my Mysophobia and OCD.''}

\textbf{Disrespect and Humiliation from Mockery.}
This harm refers to participants' feelings of being disrespected, demeaned, or embarrassed as a result of harassment.  This type of harm shows a strong connection with disability. It is often triggered by disability-centered mockery and slurs, as some participants (N=13) described being humiliated because of their disabilities. In some cases, even making jokes about disability can also make PWD feel disrespected and hurt.
For instance, P54 (male, mobility disability) mentioned, \textit{``They were stumbling around and making jokes at the expense of those who actually have disabilities. It felt incredibly disrespectful and hurtful.''}
Along with the feeling, our participants often mentioned questioning self-worth or hurting self-esteem, suggesting that this feeling could lead to further internal harm. P65 (male, attention deficit and autism) described the doubt of self-worth following with humiliation, \textit{``The mocking felt like a punch to the gut, making me feel humiliated, embarrassed, and anxious...and it made me question my self-worth and whether I was deserving of respect.''}

\textbf{Emotional Isolation due to Exclusion.}
This harm refers to the emotional experience of feeling alone, excluded, or disconnected as a result of harassment. This type of harm emphasizes the emotional loneliness that harassment triggers, rather than behavioral or social circumstances. 
Intentional exclusion can amplify feelings of abnormality and self-doubt, and participants felt they were being excluded because of their disability identity.
Some participants (N=20) reported the feeling of being lonely and isolated because of being ignored, exclusion from social activities, or ableist words in social VR, which can further lead to a strong sense of abnormality. 
\revise{P62 (female, autism) expressed the feeling of being unseen, \textit{``They looked at me and kept talking to each other like I wasn't there. I stayed for a bit, hoping they’d include me, but they didn't. I felt invisible and left out.''}}

\subsubsection{Category \#2: Internalized Harm}

Internalized harm is a form of psychological harm that extends beyond the negative emotions captured under emotional harm. It reflects deeper and more enduring impacts on an individual's physical, mental, and cognitive well-being, such as sleep problems, erosion of self-worth, dissociation, difficulty in focusing, and, in some extreme cases, a desire for self-harm. 
This harm focuses on the sustained cognitive and emotional disruptions that can interfere with daily functioning and long-term mental health problems.

\textbf{Cognitive Intrusion.}
Cognitive intrusion is the persistent presence of harassment-related thoughts, memories, or mental images that disrupt one's cognitive balance. 
Cognitive intrusion can have a more severe impact on people with disabilities. Participants described feeling trapped in the experience for an extended period (e.g., a week or even several months), with the incidents repeatedly replaying in their minds and significantly affecting their mental health.
Participants (N=31) described such disruptions both during and after the harassment. In the moment, they often felt stuck, frozen, or unable to respond to the harassment. 
This sense of situational entrapment was often linked to disability-centered insults and mockery, where participants felt trapped in the moment and struggled to respond to the demeaning words directed at their disabilities.
For example, P44 (male, autism) was overwhelmed by targeted mockery, \textit{``In the moment, I felt overwhelmed, and I had trouble thinking clearly or deciding how to respond. I felt trapped even though I could technically leave the space because harassment felt personal and targeted.''}

Cognitive intrusion also extended beyond the immediate encounter and became recurring thoughts, rumination, and difficulty processing the experience of disability-based harassment even after the harassment. P22 (female, autism) reflected on the lingering mental stress of her harassment experience,\textit{``I just kept lying there, replaying it in my head, feeling hurt and overthinking everything.''} 
Nearly half of participants who reported these harms (17 of 31) explicitly tied these harms to disability-targeted verbal harassment, such as being accused of faking a disability, insults and mockery directed at disability, or being treated as incapable because of one's disability. These findings suggest that verbal harassment not only causes distress but also disrupts PWD's cognitive processes that persist long after the harassment experience.

\textbf{Erosion of Self-Worth.}
This harm is defined as a diminished sense of self-confidence and personal value, where harassment leads to a devaluation of their own self, abilities, or experiences. 
This type of harm also has a strong connection with disability identity, which was frequently mentioned in our data, suggesting that PWD may have a higher tendency to feel unworthy due to their vulnerability. 
Some participants (N=33) reported feelings of unworthiness, self-doubt, shame, or a loss of self-esteem, often caused by ableist comments, mockery, or social exclusion. 
Rather than remaining external, demeaning language sometimes became internalized, reshaping how participants perceived themselves. 
For instance, P32 (female, blind or low vision) described how sarcastic remarks from coworkers gradually undermined her self-concept, 
 \textit{``Hearing hateful or mocking language can cause someone to internalize those messages and begin to believe that they are lesser than others...I just let people look down on me.''} P11 (male, attention deficit, autism, mental health condition and mobility-related disability) also reflected on how contempt in social VR led him to question his competence, \textit{``They just treated me with contempt, implying I was incapable of doing anything...The experience caused me to lose confidence in how I presented myself in social VR, and I second-guessed everything I did.''}

\textbf{Harmful Coping Through Substances.}
This harm builds on the definition of substance use as the continued use of alcohol, illicit drugs, or the misuse of prescription or over-the-counter medicines with negative consequences~\cite{edition2013diagnostic}. 
In our context, participants (N=9) described their reliance on alcohol or prescription medication as a way to cope with harassment in social VR, particularly when facing stress, depression, or sleep disturbances triggered by disability-related harassment.
While these coping strategies may provide short-term relief, they ultimately would hurt participants' mental health and well-being. 
For example, P9 (female, mental health condition) described how depression caused by her prior harassment experience led to a relapse with prescription painkillers, \textit{``The depression caused me to self-isolate and not speak to any loved ones/friends for a month. It also caused me to relapse on prescription painkillers. I binged on these for a week.''}

\textbf{Difficulties with Sleep.}
This harm includes having trouble falling asleep, staying asleep, or experiencing nightmares following harassment incidents. Participants (N=16) often linked these disruptions to recurring thoughts about the harassment incidents, such as replaying the incidents or recalling mocking laughter.
For example, P6 (male, health-related disability) reported experiencing nightmares in which the harasser reappeared, \textit{``The stranger kept dwelling on my mobility issue...Ongoing nightmares and insomnia. Some of the dreams are quite violent and include the stranger who was harassing me in VR.''} 
% Sleeping problems were also common as a result of sexual harassment or random ignorance.

\textbf{Dissociation and Focus Disruption.}
This harm refers to experiences of feeling detached from oneself, emotional numbness, or being unable to concentrate and maintain daily routines after harassment. 
This form of harm was less common in our data  (N=4) but nonetheless significant. 
For example, P28 (male, blind or low vision and mobility-related disability) described feeling dissociated from his own body, \textit{``I felt dissociated from my own self and sometimes felt heightened awareness of my own body which undermined my body image.''} 
P1 (male, health-related disability) wrote about having difficulty maintaining his daily routines, \textit{``I also felt emotionally numb and disconnected, which made it harder to engage in regular routines and maintain a sense of control over my personal well-being.''} 
It is worth noting that, unlike other harms that were often closely tied to participants' disability identity, dissociation and focus disruption appeared more directly connected to the traumatic experience of harassment itself rather than the disability identity.

\textbf{Self-Harm Desire.}
This harm refers to participants expressing thoughts to harm themselves.
Although this type of harm was rare in our data (P1, P15), it represents the most severe harm in this category. As P1 (male, health-related disability) shared, \textit{``At times, I used substances to cope, and I even experienced occasional thoughts of self-harm.''} Due to the limited data, we were unable to identify a clear connection between self-harm desire and disability identity.

\subsubsection{Category \#3: Social Harm}

Social harm refers to the impact of harassment on participants' social connections, relationships, and their perceived sense of safety in social environments. Unlike emotional or psychological effects, social harms extend to the ways in which harassment reshapes interpersonal interactions, erodes trust, and changes social behaviors. This harm often intersects with the trauma of harassment, which also influences how participants approach future social interaction.

\textbf{Social VR Avoidance.}
This harm refers to participants' intention to reduce or avoid social interaction in virtual social spaces after harassment. 
Participants (N=13) often mentioned being less willing to re-enter social VR apps, avoiding social rooms in VR, or completely avoiding the use of social VR for a period of time. 
Since social VR avoidance is strongly connected with traumatic experiences, participants experienced an extended period of avoidance (e.g., a few weeks or even several months) or developed a strong sense of fear about using social VR again.
Eight out of the 13 participants explicitly linked their avoidance to trauma from disability-related harassment. 
For example, P20 (female, learning disability) noted, \textit{``The experience increased my anxiety and fear about using VR Chat, and it took a long time for me to go back to VR gaming.''} 
Similarly, P23 (male, mobility-related disability) stated clearly that he needed to recover from the anxiety and being scared to re-enter the VR spaces, which highlighted how harassment directly deters PWD from participating in social VR,\textit{``I am scared to re-enter the VR spaces after the harassment and choose to avoid the spaces and first recover from the anxiety.''}

\textbf{Offline Social Withdrawal.}
This harm refers to PWD's reduced willingness to form new connections, make friends, or participate in real-world social activities after harassment in VR. 
Participants not only avoided using social VR apps but also spent less time socializing in person, isolating themselves from friends, family, and others, which severely eroded their social relationships.
Participants (N=17) described this influence, 
with 8 of them directly connecting offline social withdrawal to trauma from disability-related harassment in social VR. 
We also found a connection between social withdrawal in the real world and traumatized experience. 
For example, P45 (male, mental health condition and mobility-related disability) expressed the fear of being excluded after the harassment, \textit{``This anxiety made me reluctant to actively participate in social activities for a period of time later on, fearing being excluded again.''}  

While in many cases this social withdrawal was voluntary, we had one case in which the participant experienced externally imposed isolation. 
P14 (female, autism) described her experience of being excluded in real life after facing social exclusion related to her autism in VR, \textit{``I no longer had someone who could help me when I needed help with schoolwork and not judge my learning capacity.''}

\textbf{Disability Identity Concealment.}
This harm refers to PWD's reluctance or refusal to disclose their disability status in social activities after experiencing harassment, even though they chose to disclose their disability before. This is a new type of harm that is specific to PWD. Participants (N=4) reported their regrets about prior disclosure, being less willing to disclose in future social interactions, or decisions to never share their disability status again.  
For example, P50 (prefer not to say, blind or low vision) regretted disclosing their disability after the harassment, \textit{``If only I had not disclosed my disability, I would not be bullied.''}
In P31 (female, mobility-related disability)'s case, she decided to remove the disability signifier from her avatar after others mimicked her disability with the same disability signifiers on their avatars, \textit{``I even had to remove the avatar for a while just to avoid being mocked and to feel safe for being myself.''}
These cases highlight how harassment fosters PWD's concealment behaviors, which diminishes the openness of the social VR environment and PWD's willingness to express disability identity.

\textbf{Erosion of Social Trust.}
This harm refers to a diminished sense of security, belonging, and confidence in social VR or real-world social environments.
The erosion of social trust for PWD is related to their disability identity, i.e., participants felt unaccepted by society and viewed as abnormal or ill because of their disability.
Participants (N=18) often described feeling unsafe or mistrustful after the harassment, especially after being asked invasive questions or being mocked because of their disability identity. 
For example, P11 (male, attention deficit, autism, mental health condition, and mobility-related disability) mentioned the sense of ``mistrust'' in social VR, \textit{``They called me `crippled,' and called me out for faking disability...The harassment led me to develop a sense of mistrust and fear in social VR spaces.''} Moreover, some participants also mentioned a deep loss of social belonging, making PWD feel that they do not belong in that space. As P53 (non-binary, mental health condition) explained, \textit{``It triggered this deep sense of not belonging that stuck with me longer than I expected.''} 

In some cases, similar mistrust could extend beyond social VR to real-world social interactions. 
For example, P17 (female, multiple sclerosis) lost trust in ``people,'' which is also tied to the stereotypes about disability in real life, \textit{``I lost trust in people because of how I was treated, so it is how society views disabled people, and it is the stereotypes about disability that are existing within us.''}

\subsubsection{Category \#4: Somatic Recall Harm}

We propose somatic recall harm, a newly defined category in our framework, which refers to situations where the body ``remembers'' and reactivates symptoms in response to harassment. 
This includes triggering their prior trauma experiences and their disability-related symptoms, and forming new trauma that may become lasting memories. Unlike internalized harm, somatic recall harm specifically emphasizes the embodied activation or formation of trauma.

\textbf{Disability-Related Symptoms Reactivation.}
This harm refers to harassment that activates or worsens disability-related symptoms, such as gastritis flare-up, autistic responses, neurological issues, panic attacks, or other condition-specific symptoms experienced by PWD. This is a new type of harm that is specific to disability-related symptoms reported by our participants (N=11). As an example, P19 (non-binary, autism) described how mockery of their disability triggered autistic symptoms, \textit{``The other individuals have made funny jokes about me and how I was in VR with my disability. Some call me names because of my disability. The harassment has always made me feel tense, sweat, and experience rapid heartbeat. It also triggers some autistic symptoms.''}

Similarly, P2 (male, health-related disability and mental health condition), who has Obsessive–Compulsive Disorder (OCD) and mysophobia particularly related to his nose picking, stated the harassment worsened his symptoms, \textit{``When I shared that I have an extreme fear of people who pick their nose, they picked their nose and laughed. I said I was being serious. They said I would never be able to go outside because everyone picks their nose. This gave me extra anxiety. I kept thinking about what they said and wondered if it was true. If it was true, then literally everything has been touched by a person who picks their nose, which causes me great anxiety to the point where I have to take extreme measures just to go shopping for something like groceries.''}

\textbf{Trauma Formation and Re-Triggering.}
This harm refers to harassment incidents that either trigger past trauma experiences or create new trauma. 
The trauma is deeply tied to disability-related traumatic experiences.
Participants (N=34) reported that they often felt heightened worry, fear of recurrence in the future, or became more cautious and vigilant in similar situations. 
For example, P44 (male, autism) explained how harassment resurfaced prior painful experiences,  \textit{``The experience made me feel anxious, overwhelmed, and unwelcome. It reminded me of real-life situations where I have faced similar treatment, and it made it hard to feel safe or comfortable returning to that platform for a while.''} 

In other cases, harassment created new trauma, which further led to other types of harms, such as social withdrawal. Because of the new trauma experiences, 15 of 34 participants noted that they avoided VR or social activities to prevent re-experiencing harm.
P54 (male, mobility-related disability) explained,
\textit{``It was hard to shake off the fear of encountering the same harassment again. Every time I entered a virtual space, I found myself on edge, worrying about what might happen. I felt like I had to constantly watch my back and be cautious about who I interacted with. This anxiety made it difficult to enjoy the experience and connect with others.''}
These cases show that harassment can influence PWD's trauma responses by reactivating past pain and, at the same time, creating new trauma experiences that may persistently change PWD's social behaviors.

\subsubsection{Category \#5: Sensational Harm}
Sensational harm refers to the negative impact of harassment incidents via phantom sensations, which is a phenomenon where immersion in a VR environment causes the brain to generate physical sensations (e.g., touch, warmth, or pain) on the physical body without real, physical stimuli~\cite{deldari2023investigation}. This type of harm may blur the boundary between the virtual and real experiences. 
When harassment activates these phantom sensations, participants may experience unique, embodied harms that extend beyond psychological distress.

\textbf{Phantom Sensations.} 
Phantom sensations are unreal but vivid sensory experiences that are closely related to harassment due to one's disability. All the phantom sensations reported by our participants were relevant to their specific disabilities.
Participants (N=3) explained that these sensations would occur even though there was no physical contact when the harassment took place, such as feeling as if something was plucked from the eyes or sensing an impending seizure. 
For example, P10 (female, attention deficit, blind or low vision and mental health condition) provided an example of it, \textit{``Later in the process someone kept commenting saying my name, saying I am not able to see that because of my eyes, and the rest would laugh and agree with them... I felt as if someone had physically plucked out my eyes and made fun of them.''} 
\revise{P61 (female, neurological disorder) also described the sensation of an impending seizure triggered by flashing lights, ``With the intense flashes, I felt like I would experience a seizure the next minute. Before I could block the user or take any safety action, I was already having the feeling that a seizure was on its way.''}
Such cases illustrate that harassment in social VR \revise{may be associated with} phantom embodied experiences that feel painful and real.

\textbf{Embodied Reactions.}  
This harm refers to the physical responses and discomfort participants experienced during or after harassment in VR, ranging from temporary sensations to more intense pain. Participants (N=26) reported reactions such as chest tightening, chills, sweating, trembling, racing heartbeat, nausea, fatigue, or other bodily discomforts and pain. While this harm is not always tied directly to disability, participants' reactions may be further intensified by the unique immersive qualities in VR. 
\revise{It is possible that} the immersive and embodied nature of VR can blur the boundary between the virtual and the real, making PWD feel as if harassment in VR were happening in reality, thereby triggering strong physical reactions. For example, P64 (female, speech-related disability) described, \textit{``I felt hot on my head and face because I was so embarrassed by the incident.''} Similarly, P3 (female, mental health condition) shared, \textit{``I started to feel my heart racing and started to feel very warm, like I was embarrassed, but I shouldn't have been.''} 
In other cases, participants reported more painful experiences, such as headaches or chest pain. As P50 (prefer not to say, blind or low vision) recalled, \textit{``When that person ignored me after I disclosed my disability, I felt a tight pain in my chest. It literally hurts to be ignored,''} further demonstrating that harassment in VR could produce tangible sensations of pain and discomfort.

\subsection{Severity Perception of Harms}
\begin{figure}
  \centering
  \includegraphics[
  width=\linewidth,
  alt={A heatmap table summarizing severity ratings for 19 harm subtypes. Each row represents a harm subtype, and each column reports the number of ratings, minimum score, maximum score, mean severity rating, and variance. The harms are sorted by mean severity. Cells are color-coded according to severity values. Disability identity concealment has the highest mean severity rating (4.25), followed by phantom sensations (4.00) and social VR avoidance (3.92). Disrespect and humiliation from mockery have the lowest mean rating (2.92).}
]{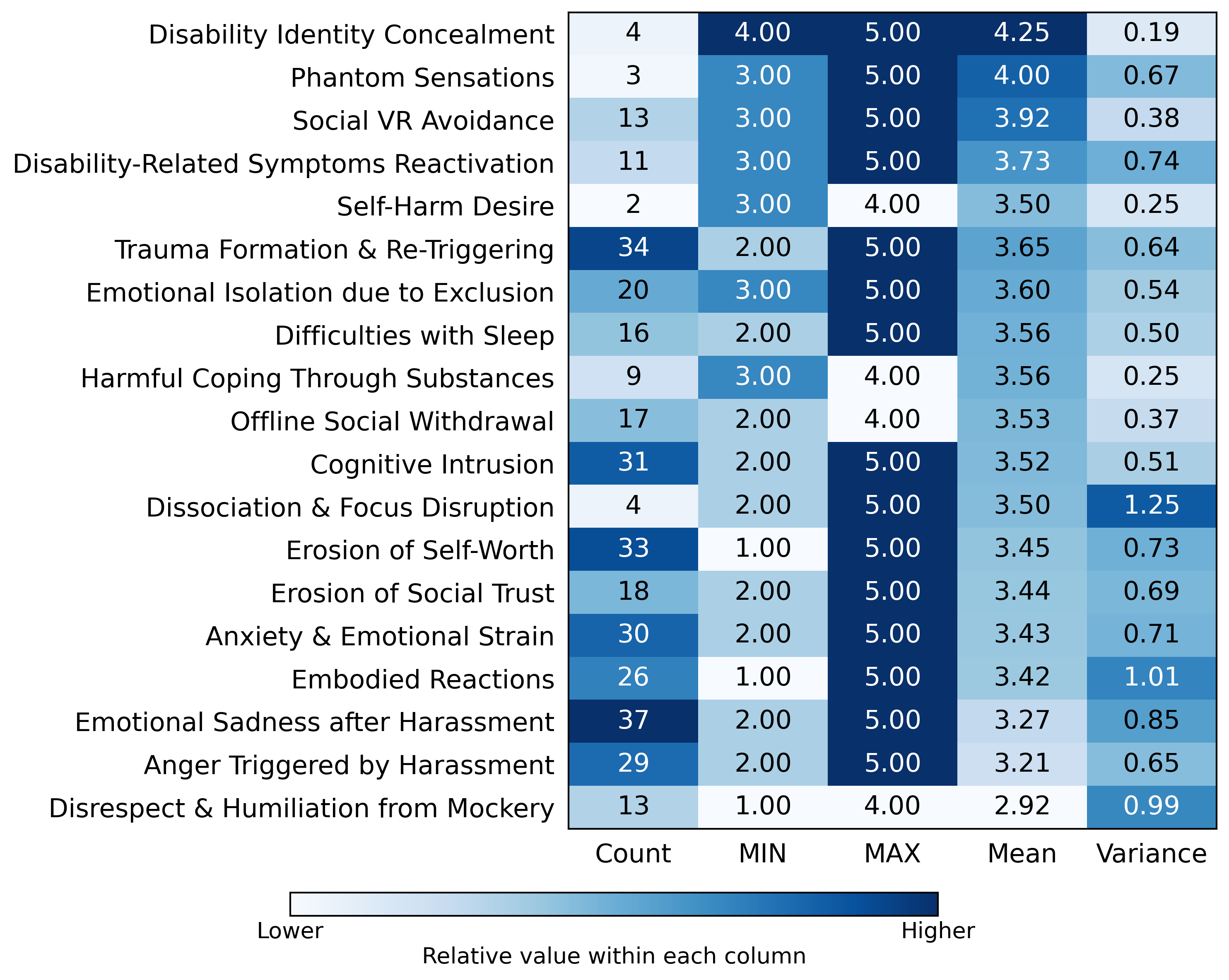}
  \caption{Severity ratings across harm sub-types, showing each sub-type's rating count, minimum, maximum, mean, and variance, sorted by mean rating.}

\Description{A heatmap table summarizing severity ratings for 19 harm subtypes. Each row represents a harm subtype, and each column reports the number of ratings, minimum score, maximum score, mean severity rating, and variance. The harms are sorted by mean severity. Cells are color-coded according to severity values. Disability identity concealment has the highest mean severity rating (4.25), followed by phantom sensations (4.00) and social VR avoidance (3.92). Disrespect and humiliation from mockery have the lowest mean rating (2.92).}
  \label{fig:vis_severity}
\end{figure}

To understand the perceived severity of each harm type, we analyzed the severity ratings and visualized the results in Figure~\ref{fig:vis_severity}. For the counting of harm types, ``Emotional Sadness after Harassment,'' ``Trauma Formation and Re-Triggering,'' and ``Erosion of Social Trust'' had higher occurrences, indicating they were more common across all the harm types. For mean ratings, ``Disability Identity Concealment'' had the highest rating, which might be because it was about disability identity disclosure and affected both online and real-world social experiences deeply. ``Phantom Sensations'' also had a higher rating, mostly because it was related to specific disability symptoms. Interestingly, the rating of ``Disrespect and Humiliation from Mockery'' was relatively low, and participants described it as \textit{``only affecting me for a short time''} and \textit{``it did not have an effect on my daily routine.''} In addition, ``Dissociation and Focus Disruption'' and ``Embodied Reactions'' had high variation, showing that participants' opinions varied.
We observed that higher ratings were mostly related to participants' descriptions in their open-ended responses such as \textit{``hard to get over''} and \textit{``affects social ability''}, 
\revise{and other participants mentioned turning to ``seek a therapist for help'' or ``seek support from people close to me''}
; while lower ratings were related to comments such as \textit{``affects a short time''} and \textit{``can cope with it myself.''} 

In summary, when harms extended to the real world and were closely connected to disability identity, the perceived severity was higher, and when the harm lasted a short time and did not affect daily life, the rating was lower.

\section{Discussion}
\label{sec:discussion}

In this paper, we presented 19 types of harms in five high-level categories. In this section, we first draw on Critical Disability Theory to examine harm, then compare the types of harms that PWD experienced across online spaces and the physical world to explore the unique harms in the social VR context, and finally derive design implications for developing safer social VR environments for PWD.

\new{
\subsection{Rethinking the Harms in Social VR: A Critical Disability Theory Perspective}
To examine the harms experienced by PWD in social VR, Critical Disability Theory (CDT) ~\cite{sep-disability-critical} provides a vital lens for understanding these harms. By applying CDT to our taxonomy of harms, we can see how platform designs and social dynamics actively marginalize users who have disabilities.
}

\new{
In our category of social harm, specifically disability identity concealment, CDT helps contextualize this harm as a forced reaction to the culture of the ``normate.'' Garland-Thomson~\cite{thomson_extraordinary_2017} defines the normate as the \textit{``constructed identity of those who, by way of the bodily configurations and cultural capital they assume, can step into a position of authority and wield the power it grants them.''} In social VR, disability-centered harassment reinforces these norms by punishing bodies, behaviors, and ways of interacting that are seen as different. As a result, PWD are not just hiding their identities to avoid individual harassers. Instead, they hide them as a way to protect themselves in a hostile social environment that often devalues disability.
}

\new{
A core premise of CDT is that disability is not located solely within individuals, but is produced by the ``rigidity, faultiness, deficits, and pathological structures'' embedded in social environments~\cite{linton_what_2005}. Our findings show that harassment in social VR can lead some PWD to avoid VR and even withdraw from offline social activities. This suggests that VR platforms themselves can act as disabling environments. When platform design, reporting systems, and moderation fail to stop ableist harassment or create safe spaces, they can push PWD out of these digital spaces. The harm, therefore, goes beyond the immediate experience of harassment. It can exclude PWD from online social life, weaken their trust in social technologies, and reinforce feelings of marginalization.
}

\subsection{Unique Harm in Social VR}

\new{Some of the harms identified in our study (e.g., anxiety, sadness, and social withdrawal) have also been reported in prior online harassment research. However, our findings suggest that their causes, expressions, and impacts are distinct in social VR, possibly due to its embodied, immersive, and interactive nature. For example, emotional harms in our data, such as feelings of disrespect and isolation, feel closer to real-world experiences and appear more complex than those typically described in online settings. In addition, we observe that social withdrawal not only stays within the virtual platform but also often extends into offline social life, which is a pattern that is rarely discussed. These observations point to broader differences in how harm is experienced in social VR. We examine these differences in relation to prior online and real-world contexts below.}

Prior work suggested that PWD may receive various ableist comments (hateful words and ableist microaggressions), disability-centered insults, accusations of faking disability, and even direct threats on social media~\cite{heung2024vulnerable}. As a result, they reported feelings of anxiety, sadness, isolation, and anger after receiving such disability-centered harmful comments~\cite{heung2024vulnerable}. In fact, PWD may also suffer from similar feelings after experiencing cyberbullying, cyberharassment, or other cybercrimes targeting people with disabilities~\cite{chadwick2019online, triantafyllopoulou2025safer, baltzar2024forced, normand2016cybervictimization}.  
These harms include feelings of depression, anxiety, frustration, and anger, which were also reported in our emotional harms in the social VR context. 

Uniquely, in our data, many participants described feeling disrespected, humiliated, and isolated because they were targeted for their disability. These emotional details are rarely discussed in previous studies on harassment experiences of PWD in online environments. Interestingly, participants also mentioned similar experiences in the physical world, such as feeling embarrassed, humiliated, isolated, and exposed~\cite{equality2011hidden, carr2017mental}. This \revise{may suggest} that experiences in social VR are more closely connected to those in real life, highlighting how the embodied and immersive nature of VR can evoke real-world emotional reactions while still sharing some similarities with traditional online spaces.

Similarly, prior work showed that PWD's harassment experiences may cause trauma, such as fear of posting online ~\cite{heung2024vulnerable, alhaboby2016language} or the triggering of past traumatic memories~\cite{equality2011hidden, lund2021disability}, which often leads to deliberate withdrawal from or hiding their disability identity as a means of protection~\cite{arapol2023uncovering,alhaboby2016language, sin2009disabled}.
These harms echo what PWD have experienced in social VR, as we reported in this paper. However, these studies rarely mentioned the social withdrawal extending from the online space to the offline world. Some work reported that in online spaces, participants become socially isolated and excluded from safe online engagement~\cite{triantafyllopoulou2025safer, baltzar2024forced}. In our study, participants not only avoided using social VR, often staying away from VR platforms for weeks, but also showed a reduced willingness to engage in real-world social activities after experiencing harassment in VR. This \revise{may reflect} a deeper social impact that extended beyond virtual spaces and affected their real-world socialization. Similar behaviors can also be found in the aftermath of harassment experienced by PWD in the physical world, where victims chose to move to a new house, leave college, or go out less to avoid being targeted by neighbors~\cite{sin2009disabled, equality2011hidden}.

In reality, the harms that PWD experienced in the physical world may result in severe or even extreme consequences. For example, in a prior case, the victim formed long-term trauma and remained indoors for over a year without stepping outside of the house~\cite{equality2011hidden}. Fortunately, we have not yet observed such cases in social VR. However, the overlapping harm experiences between social VR and the physical world, together with the common experiences between social VR and traditional online spaces, point to a concerning possibility---to some extent, social VR is blurring the boundary between the virtual world and the physical world. While this mixed, immersive experience \revise{could allow users to experience both worlds, it may also lead to the embodiment of harassment and the associated harms from both worlds.} In other words, while social VR combines the best of both worlds, it also suffers from the dark side of both worlds in terms of harm. The consequences of these harms can be further amplified for PWD because of their disability-centered vulnerabilities. 
As such, researchers and practitioners need to work together diligently to preserve the immersive experiences in social VR while protecting PWD from the possible harms so that they can enjoy the benefits introduced by these new technologies.

In contrast, compared to the traditional online spaces and physical world, we did not observe cases in which PWD reported financial harms or physical harms. As a comparison, this type of harm is rather common on social media. For example, prior work~\cite{heung2024vulnerable, triantafyllopoulou2025safer, carr2017mental} has
found that PWD suffered long-term financial harm, such as reduced opportunities for financial compensation, visibility, and hindered educational or employment prospects. Moreover, ableist content on social media can negatively influence public perceptions of PWD, making it harder for them to seek financial support opportunities. For physical harm, individuals reported experiencing physical injuries resulting from physical abuse~\cite{carr2017mental, viluckiene2015relationship}. It can also manifest as aggressive behavior as a consequence of targeted violence, including damage to their own property~\cite{sin2009disabled}. Luckily, none of these issues showed up in our data.

We believe that, even though we did not observe financial harms or physical harms among our participants, we still need extra precautions to cope with similar harms should they migrate to social VR. In the next section, we will offer design implications based on our findings.

\subsection{Design Implications}

Based on our results, we make the following design implications to mitigate the harms caused by disability-related harassment in social VR. \revise{These implications should be interpreted within the context of our participants' disability groups.}

\subsubsection{Prevent the Harm by Using AI-based Detection}
To mitigate the harm resulting from disability-centered harassment, the obvious approach is to reduce harassment incidents and stop the harm at its root cause. 
In our results, 
verbal forms of harassment (e.g., disability-centered insults and mockery, treating disability as inability, and accusing of faking disability) were the most common (N=81) in our data. Other harassment types, such as intentionally triggering disability symptoms or mimicking disability, can involve complex forms of harassment that combine both verbal and environmental elements. Although these types were less frequent in our data, they specifically targeted certain kinds of disabilities to intentionally harass or trigger disability symptoms, causing severe harm to PWD. In summary, implementing AI-detection systems can be highly effective in mitigating the harms associated with these types of harassment by preventing them from happening in advance.

For example, AI-based detection and filtering of ableist or demeaning language could play an important role in reducing the occurrence of emotional harm. Previous AI moderation and filtering systems have primarily focused on general VR harassment~\cite{ schulenberg2023towards,xu2024safe, wang2024hardenvr}, behavioral content detection~\cite{weerasinghe2025beyond, fiani2023big}, or consent-based mechanisms~\cite{connor2023consensual} to prevent harassment, with insufficient attention to disability-centered slurs, ableist terms, or mockery. We suggest that platforms incorporate AI detection for both voice and chat channels with a specific focus on disability-centered hate words and slurs (e.g., crippled, handicapped), filtering demeaning language before it can be seen or heard by other users. Heung et al.~\cite{heung2025ignorance} focused on designing AI moderation systems to address ableist hate, but their work is limited to social media and lacks exploration of more diverse AI moderation approaches in complex environments such as social VR. We propose designing more advanced and diverse AI moderation mechanisms that monitor not only voice and text interactions but also avatar interactions and environmental object placements. Such systems can proactively detect and prevent inappropriate behaviors between avatars or the creation of spaces and objects that may cause discomfort to others, enabling more comprehensive prevention of harassment in social VR.
\revise{Although promising, we should also be careful about the risks of using AI in such features. For example, AI-based detection may result in false positives and unreliable outcomes~\cite{erol2025cana, mugiraneza2025challenges}. For AI-driven mental health support, there may be risks such as generating biased or harmful responses and leaking patients' personal data in mental health interventions~\cite{lawrence2024opportunities}. Thus, any AI-based solutions should be carefully designed and thoroughly tested before being broadly implemented.
}

\revise{
\subsubsection{New Features in Social VR Platform to Mitigate Harms}
Current social VR platforms leave substantial room for features that mitigate harm during harassment or block its sources before harm occurs. Prior work on social media has explored real-time detection of sensitive content to support users with misophonia and prevent the formation of trauma~\cite{ammari2026remote, chen2022trauma}. Social VR could similarly integrate mechanisms that automatically detect sensitive content (e.g., images and videos displayed in a room) that may trigger symptoms of disabilities such as misophonia or neurological disorders. To further mitigate somatic recall triggered by such stimuli, platforms could explore ways to rapidly block the sources (e.g., bright flashes or other triggering imagery). Platforms could also offer a safe space: a quick-access button that lets users immediately leave the current room and enter a private, protected environment. This safe space should additionally provide aftercare support to prevent more severe and lasting harm, such as internalized harm.
}

\subsubsection{Educational Nudges to Raise Awareness of Harms}
We believe that designing educational nudges about harms can also be a useful intervention, as it raises other users' (including the harassers') awareness of the potential harm they may bring to PWD. In our survey, some participants described certain harassment as microaggressions, noting that people may not realize their words or behaviors can cause internalized harm. Although intended as jokes, such actions can still cause harm to PWD. Therefore, it is important to raise awareness of the potential harm caused by harassment, either intentional or unintentional. Prior works have explored the design of educational nudges, consent mechanics, and boundary-setting cues to prevent harassment and promote safer interactions in social VR~\cite{liao2025building, schulenberg2023we, connor2023consensual}, yet these studies primarily focus on general user safety and do not specifically address the unique harms experienced by PWD or how nudging mechanisms can be tailored to mitigate such disability-related vulnerabilities. Platforms could display these nudges in public social VR rooms to inform users that certain behaviors may cause harm, or automatically present the nudges when users enter these spaces to educate them about treating other users in a respectful manner.

\subsubsection{AI Therapists after Harm Has Occurred}

Even if we can design various safety mechanisms, we cannot completely prevent harassment from happening. Some participants mentioned in our survey that they faced significant challenges in overcoming the trauma resulting from harassment. Other participants 
mentioned that they turned to a therapist for help or sought support from close ones, which could ease the negative impacts.

Inspired by these results, we believe that offering mental health support services to mitigate the harms after harassment can be an effective way to address its aftermath. 
Recent studies have also explored the use of virtual human agents and immersive systems for mental health support in VR settings~\cite{wei2025virtual, wang2024evaluating, sharma2023human}. Building on these findings, we propose that social VR platforms can incorporate virtual agents, either human-operated or AI-driven, to provide immediate assistance, consultation, and post-trauma psychological support for victims of harassment. These agents should especially focus on providing mental health support services for people with disabilities. The AI systems should be trained to interact with PWD carefully, using respectful and appropriate language, and should be developed with datasets that include various forms of harassment targeted at PWD as well as disability-related symptoms to ensure more sensitive and informed support.
In addition, creating dedicated social VR rooms or communities where users can share their experiences and support one another in recovering from pain and trauma may also be an effective approach.

\subsection{Limitations}
Our study has several limitations. First, we recruited 67 participants exclusively from the United States. The results may not
be applicable in other cultural contexts or countries. Second, although we aimed to recruit participants with diverse disabilities, our sample was concentrated among participants with mental and mobility-related disabilities, and we had limited representation of less common disability types. Third, the survey format constrained the level of detail we could collect about harassment incidents. Some participants provided only brief descriptions, which may have led us to miss nuances of potential harms that were not fully described. \revise{We chose the survey method instead of interviews, as our goal was to identify as many types of harm as possible. Future work could use interview studies to develop a deeper understanding of PWD's harm experiences.}

\section{Conclusion}
\label{sec:conclusion}
We presented a harm framework centered on the experiences of PWD in social VR, identifying five categories of harm and 19 subtypes that detail how harassment leads to negative impacts. Through a systematic literature review and a survey with 67 PWD, we further analyzed severity ratings.
Finally, we discussed our findings from the critical disability theory perspective and revealed unique harms in social VR. We also provided design implications for mitigating the harms after or during harassment in social VR.

\begin{acks}

We are grateful to all anonymous participants and reviewers for their time and thoughtful contributions to this research.
This work was supported in part by the National Science Foundation under Grant No. IIS-2328182, No. IIS-2608524, and No. CNS-2442221, and a Google PSS Faculty Award. Any opinions, findings, conclusions, or recommendations expressed in this material are those of the authors and do not necessarily reflect the views of the sponsors.

\end{acks}

%%% Local Variables:
%%% mode: latex
%%% TeX-master: "main"
%%% End:

\bibliographystyle{ACM-Reference-Format}
\bibliography{reference}

\clearpage
% % --- Appendix ---%
\appendix
\section{Appendix}
\label{tab:appendix}

\subsection{Harassment Definitions}
\label{tab:harassment-definitions}

\vspace{0.5\baselineskip}

\begin{center}
\begin{minipage}{\textwidth}
\centering
\footnotesize
\renewcommand{\arraystretch}{1.2}
% \tagpdfsetup{table/header-rows={1}}
\begin{tabular}{@{}p{3cm}p{6cm}p{7cm}@{}}
\toprule
\textbf{Harassment Type}
&
\textbf{Definition}
&
\textbf{Examples}
\\
\midrule

Disability-Centered Insults and Mockery & The use of slurs, insults, or mocking behaviors in social VR that target a person’s disability, including subtle to explicit actions that ridicule, devalue, or demean the individual based on their condition.~\cite{equality2011hidden, healy2020spreads, holzbauer2010typology, holzbauer2008disability,lindsay2023ableism,sin2009disabled, heung2024vulnerable, alhaboby2016language, sannon2023disability, zhang2023diary} & 
e.g., using terms like ``cripple'' or ``handicapped'' in a hurtful way, mocking someone’s disability, or making sarcastic comments like ``oh, you can't hear, haha'' \\
\midrule
Treating Disability as Inability & The assumption that having a disability means a person is less capable, unable to participate in daily life or society, or childlike in social VR.~\cite{equality2011hidden, holzbauer2008disability, lindsay2023ableism, hackett2020disability, lindsay2022time, timmons2024ableism, heung2022nothing, zhang2023diary,angerbauer2024part} & e.g., treating disability as inferior, speaking in a babyish tone, exaggerating normal actions as ``inspiring,'' or making comments that suggest they can’t contribute or be independent \\
\midrule
Sexual Harassment & Involves sexually suggestive actions or conversations that can feel invasive or threatening in social VR, especially when done without consent.~\cite{equality2011hidden, byrne2018prevalence, heung2024vulnerable,sannon2023disability, schoenebeck2023online, blackwell2019harassment, hinduja2024metaverse, abhinaya2024enabling} & e.g., making sexual gestures toward others, sending or speaking sexual messages, jokes, or unwanted flirtation \\
\midrule
Physical Harassment & Involves aggressive, unwanted, or non-consensual physical interactions in social VR that target people with disabilities.~\cite{equality2011hidden, byrne2018prevalence, sin2009disabled, zhang2023diary, freeman2022disturbing, blackwell2019harassment} & e.g., punching a disabled user’s avatar, throwing virtual objects at them, forcing avatars out of wheelchairs, standing too close, or interfering with assistive technology without permission. \\
\midrule
Social Exclusion & Happens when people with disabilities are deliberately ignored, left out of conversations or group activities, or treated as unwelcome in social VR.~\cite{equality2011hidden, holzbauer2010typology, lindsay2023ableism, hackett2020disability, lindsay2022time, sin2009disabled, heung2022nothing, alhaboby2016language, schoenebeck2023online, zhang2023diary} & e.g., being excluded from conversations or activities, ignored during interactions, being forced to quit the game, or ``ghosted'' in social VR spaces because of their disability \\
\midrule
Mimicking Disability & Intentionally imitating disability-related traits in social VR to mock, ridicule, or entertain others.~\cite{holzbauer2008disability, zhang2023diary, schulenberg2024does, oguine2023you} & e.g., switching to an avatar in a wheelchair or using a cane to imitate a disabled user, sarcastically mimicking disability-related behaviors. \\
\midrule
Accusing of Faking Disability & Harassers in social VR deny or question the legitimacy of a person’s disability, suggesting that the individual is exaggerating, fabricating, or misrepresenting their condition.~\cite{lindsay2023ableism, sin2009disabled, heung2024vulnerable, heung2022nothing, alhaboby2016language, nario2019ableism, angerbauer2024part} & e.g., saying someone is lying about their disability, suggesting they are exaggerating their condition, or comparing them unfairly to others with different disabilities to invalidate their experience \\
\midrule
Invasive Personal Questions & Involves asking, or persistently asking, inappropriate or overly personal questions that feel invasive and uncomfortable to people with disabilities in social VR.~\cite{equality2011hidden, lindsay2023ableism, heung2022nothing, schoenebeck2023online, nario2019ableism, schulenberg2024does, chen2025understanding} & e.g., asking what happened to their body, how they became disabled, or intrusive questions about their sexual health or activity \\
\midrule
\textbf{Intentionally Triggering Disability Symptoms} & Refers to deliberate actions in social VR, following disability disclosure, where harassers intentionally engage in behaviors or manipulate the environment to provoke the specific symptoms of a person’s disability & e.g., displaying flashing lights deliberately to trigger neurological disability \\
\midrule
\textbf{Indirect Harassment} & When people with disabilities witness or overhear harassment in social VR, such as insults, mocking, or mimicking, which can still cause emotional distress or trauma even without direct engagement. & e.g., overhearing insults about disability by others or witnessing others mimicking disability by using avatars\\

\bottomrule
\end{tabular}

\vspace{4pt}
\captionof{table}{Harassment type definitions. The first eight types are drawn from prior work; bolded types are newly defined in this study.}
\label{tab:harassment_def}

\end{minipage}
\end{center}

\clearpage

\subsection{Survey Protocol}
\label{tab:survey_protocol}

\textbf{Demographics}

\begin{enumerate}
  \item What is your age group?
    \begin{itemize}
      \item 18 to 24
      \item 25 to 34
      \item 35 to 44
      \item 45 to 54
      \item 55 to 64
      \item 65 or over
    \end{itemize}

  \item What types of disabilities have you been diagnosed with? If you have more than one disability, please select all that apply:
    \begin{itemize}
      \item Attention deficit
      \item Autism
      \item Blind or low vision
      \item Deaf or Hard of Hearing
      \item Health-related disability
      \item Learning disability
      \item Mental health condition
      \item Mobility-related disability
      \item Speech-related disability
      \item Others, please specify: \_\_\_\_
    \end{itemize}

  \item How do you identify your gender?
    \begin{itemize}
      \item Male
      \item Female
      \item Non-Binary
      \item Others, please specify: \_\_\_\_
      \item Prefer not to say
    \end{itemize}

  \item What’s your ethnicity?
    \begin{itemize}
      \item American Indian or Alaska Native
      \item Asian
      \item Black or African American
      \item Hispanic or Latino
      \item Native Hawaiian or Other Pacific Islander
      \item White
      \item Prefer not to say
    \end{itemize}
\end{enumerate}

\textbf{Social VR Experience}

\begin{enumerate}
  \item What social VR applications have you used? Please select all that apply:
  
    \begin{itemize}
    
      \item VRChat
      \item RecRoom
      \item Horizon Worlds
      \item AltspaceVR
      \item BigScreen
      \item NeosVR
      \item Spatial
      \item Mozilla Hubs
      \item Cluster
      \item Roblox
      \item Sandbox
      \item Sansar
      \item Others, please specify: \_\_\_
    \end{itemize}

  \item How long have you been using social VR?
    \begin{itemize}
      \item Less than a month
      \item 1--6 months
      \item 7--12 months
      \item 1--2 years
      \item More than 2 years
    \end{itemize}

  \item How often do you use social VR?
    \begin{itemize}
      \item Very rarely: I use it less than once a month
      \item Rarely: I use it about once a month
      \item Occasionally: I use it several times a month
      \item Frequently: I use it weekly
      \item Very frequently: I use it daily or almost every day
    \end{itemize}

  \item Which of the following methods have you used to disclose your disability in social VR? Select all that apply.
    \begin{itemize}
      \item Through avatar appearance (e.g., wheelchair, prosthetics)
      \item Profile information or bio
      \item Floating name tag or status
      \item Verbal conversation with others
      \item I have not disclosed my disability
      \item Other, please specify: \_\_\_\_
    \end{itemize}
\end{enumerate}

\textbf{VR Harassment Experience}

\begin{enumerate}
  \item How would you define harassment in social VR (e.g., what types of behaviors or actions do you consider harassment)?

  \item Which of the following types of harassment have you experienced in social VR? Please select all that apply.

        \textbf{Ableist Hate Language:} The use of slurs, insults, or mocking behavior that targets people with disabilities in social VR, treating their condition as something to ridicule or devalue. (e.g., using terms like “cripple” or “handicapped” in a hurtful way, teasing someone about their disability, asking them to do things they physically can’t, or making sarcastic comments like “oh, you can’t hear, haha”)
  
      \text{Treating Disability as Inability:} The assumption that having a disability means a person is less capable, unable to participate in daily life or society, or childlike in social VR. (e.g., questioning a disabled person’s intelligence, denying their achievements in games, speaking in a babyish tone, exaggerating normal actions as “inspiring,” or making comments that suggest they can’t contribute or be independent)
      
      \textbf{Sexual Harassment:} Involves sexually suggestive actions or conversations that can feel invasive or threatening in social VR, especially when done without consent. (e.g., making sexual gestures toward others, sending or speaking sexual messages, jokes, or unwanted flirtation)
      
      \textbf{Physical Harassment:} Involves aggressive or unwanted physical interactions in social VR, especially targeting people with disabilities. (e.g., punching, slapping, or kicking a disabled user’s avatar; throwing virtual objects at them; forcing avatars out of wheelchairs; standing too close; or interfering with assistive technology without permission, including offering unwanted “help”)
      
      \textbf{Social Exclusion:} Happens when people with disabilities are deliberately ignored, left out of conversations or group activities, or treated as unwelcome in social VR. (e.g., being excluded from multiplayer games, ignored during interactions, or “ghosted” in social VR spaces because of their disability)
      
      \textbf{Mimicking Disability:} Involves users intentionally changing their avatars to imitate disabilities they do not have, often to mock or entertain others in social VR. (e.g., switching to an avatar in a wheelchair or using a cane to imitate a disabled user, leading others to laugh or treat it as a joke)
      
      \textbf{Accusing of Faking Disability:} Occurs when someone questions or denies the legitimacy of a person’s disability in social VR, often implying they are pretending for attention or benefit. (e.g., saying someone is lying about their disability, suggesting they are exaggerating their condition, or comparing them unfairly to others with different disabilities to invalidate their experience)
      
      \textbf{Invasive Personal Questions:} Involves asking inappropriate or overly personal questions that make people with disabilities feel uncomfortable or disrespected in social VR. (e.g., asking what happened to their body, how they became disabled, or intrusive questions about their sexual health or activity)
      
      \textbf{Other,} please specify: \_\_\_\_

  \item How often do you experience such harassment in social VR because of your disability?
    \begin{itemize}
      \item Very rarely: I experience it less than once a month
      \item Rarely: I experience it about once a month
      \item Occasionally: I experience it several times a month
      \item Frequently: I experience it weekly
      \item Very frequently: I experience it daily or almost every day
    \end{itemize}

  \item Where did the incident occur? (Select all that apply)
    \begin{itemize}
      \item Public VR space (e.g., open world or shared lobby)
      \item Private VR space (e.g., invite-only room or private event)
      \item I’m not sure
      \item Other, please specify: \_\_\_\_
    \end{itemize}

  \item Which platform were you using when it happened?

  \item How many people were involved in the incident?
    \begin{itemize}
      \item Only one
      \item More than one
      \item I don’t know
    \end{itemize}

  \item How would you describe your relationship to the person or people involved? (Select the option that best describes them)
    \begin{itemize}
      \item Stranger
      \item Acquaintance
      \item Friend
      \item Family member
      \item I don’t know
      \item Other, please specify: \_\_\_\_
    \end{itemize}

  \item Please describe what happened (Describe the incident in your own words).

  \item Did that ableist hate harassment experience in social VR affect you physically? (e.g., bodily injury, property damage)
    \begin{itemize}
      \item No, it has not affected me physically.
      \item Yes.
    \end{itemize}

\textit{If you selected ``Yes'' to the previous question, answer the following:}

\begin{enumerate}
  \item Which of the following physical effects did you experience? Select all that apply.
    \begin{itemize}
      \item Bodily injury
      \item Property damage
      \item Other, please specify: \_\_\_\_
    \end{itemize}

  \item Please describe the physical effects you experienced in more detail. You may include any context you feel is important.

  \item Please rate the severity of the effect:
    \begin{itemize}
      \item Not severe at all
      \item Slightly severe
      \item Moderately severe
      \item Severe
      \item Very severe
    \end{itemize}

  \item Why did you give this rating? Please briefly explain your reasoning (e.g., how long the effects lasted and how it affected your daily life, health, or feelings).
\end{enumerate}

  \item Did that ableist hate language harassment experience in social VR cause you to experience phantom sensations?
    \begin{itemize}
      \item No, I did not experience any physical sensations.
      \item Yes.
    \end{itemize}

\textit{If you selected ``Yes'' to the previous question, please answer the following:}

\begin{enumerate}
  \item What type of phantom sense did you experience?
    \begin{itemize}
      \item Touch (e.g., stroking or tapping)
      \item Warmth or heat
      \item Pain
      \item Tingling or itchiness
      \item Cold sensation
      \item Smell
      \item Other, please specify: \_\_\_\_
    \end{itemize}

  \item Please describe how it felt (e.g., how intense it was, how real it felt, or how it affected you).

  \item Please rate the severity of the effect (same severity scale as above).

  \item Why did you give this rating? Please briefly explain your reasoning.
\end{enumerate}

  \item Did that ableist hate language harassment experience in social VR affect you psychologically? (e.g., feeling depressed, anxious, or having sleeping issues)
    \begin{itemize}
      \item No, it has not affected me psychologically.
      \item Yes.
    \end{itemize}

\textit{If you selected ``Yes'' to the previous question, answer the following:}

\begin{enumerate}
  \item Which of the following psychological effects did you experience? Select all that apply.
    \begin{itemize}
      \item Depression
      \item Anger
      \item Mania
      \item Anxiety
      \item Somatic Symptoms
      \item Suicidal Ideation
      \item Psychosis
      \item Sleep Problems
      \item Memory Problems
      \item Repetitive Thoughts and Behaviors
      \item Dissociation
      \item Personality Functioning Difficulties
      \item Substance Use
      \item Other, please specify: \_\_\_\_
    \end{itemize}

  \item Please describe the psychological effects you experienced in more detail.

  \item Please rate the severity of the effect (same severity scale).

  \item Why did you give this rating? Please briefly explain your reasoning.
\end{enumerate}

\item Did that ableist hate language harassment experience in social VR affect you economically? (e.g., financial loss)
\begin{itemize}
  \item No, it has not affected me economically.
  \item Yes.
\end{itemize}

\textit{If you selected ``Yes'' to the previous question, answer the following:}

\begin{enumerate}
  \item Which of the following economic effects did you experience? Select all that apply.
    \begin{itemize}
      \item Scam (e.g. phishing)
      \item Account hack
      \item Digital asset loss
      \item Bribery
      \item Blackmail
      \item Other, please specify: \_\_\_\_
    \end{itemize}

  \item Please describe the economic effects you experienced in more detail.

  \item Please rate the severity of the effect: (same severity scale)

  \item Why did you give this rating? Please briefly explain your reasoning.
\end{enumerate}

\item Did that ableist hate harassment experience in social VR cause you any other types of harm?
\begin{itemize}
  \item No, I did not experience any other types of harm.
  \item Yes.
\end{itemize}

\textit{If you selected ``Yes'' to the previous question, answer the following:}

\begin{enumerate}
  \item Please describe the type of harm you experienced and how it affected you.

  \item Please rate the severity of the effect (same severity scale).

  \item Why did you give this rating? Please briefly explain your reasoning.
\end{enumerate}

  \item Did you feel that the harassment experience in social VR was consistent with the real-world experience?
    \begin{itemize}
      \item Very similar
      \item Somewhat similar
      \item Neutral
      \item Somewhat different
      \item Very different
    \end{itemize}

  \item Can you explain why you feel that way?

  \item Did you feel that the harm of the virtual harassment experience was consistent with the real-world experience?
    \begin{itemize}
      \item Very similar
      \item Somewhat similar
      \item Neutral
      \item Somewhat different
      \item Very different
    \end{itemize}

  \item Can you explain why you feel that way?
\end{enumerate}

\section*{Other Types of Harassment}
Repeat the same structure of questions as above for other types (e.g., Treating Disability as Inability, Sexual Harassment, etc.).

% The survey protocol can be accessed through the Open Science Framework\footnote{\url{https://osf.io/8xb27/overview?view_only=7e273d728f5b49738d2a1291ca729bf9}}.

% \input{table/harassment-fre}

\end{document}